\documentclass[final,3p,times,twocolumn,authoryear]{elsarticle}

\usepackage{amssymb}
\usepackage{amsmath}
\usepackage{array}
\usepackage[
  colorlinks,
]{hyperref}
\usepackage{lineno}
\usepackage{mdframed}
\usepackage{url}
\usepackage[x11names]{xcolor}
\usepackage{xspace}
\usepackage{xstring}
\usepackage{ifthen}
\usepackage[normalem]{ulem} 
\usepackage{xcolor,colortbl}
\usepackage{hyperref}

\newboolean{showedits}
\setboolean{showedits}{false} 
\ifthenelse{\boolean{showedits}}
{
	\newcommand{\ins}[1]{\textcolor{blue}{\uline{#1}}} 
	\newcommand{\del}[1]{\textcolor{red}{\sout{#1}}} 
	\newcommand{\chg}[2]{\textcolor{red}{\sout{#1}}{\ra}\textcolor{blue}{\uline{#2}}} 
	\newcommand{\nbe}[3]{
		{\colorbox{#3}{\bfseries\sffamily\scriptsize\textcolor{white}{#1}}}
		{\textcolor{#3}{\sf\small$\blacktriangleright$\textit{#2}$\blacktriangleleft$}}}
}{
	\newcommand{\ins}[1]{#1} 
	\newcommand{\del}[1]{} 
	\newcommand{\chg}[2]{#2}
	\newcommand{\nbe}[3]{}
}

\newboolean{showcomments}
\ifthenelse{\boolean{showcomments}}
 {
 	\newcommand{\nbc}[3]{
 		{\colorbox{#3}{\bfseries\sffamily\scriptsize\textcolor{white}{#1}}}
		{\textcolor{#3}{\sf\small$\blacktriangleright$\textit{#2}$\blacktriangleleft$}}}
	
 }{
 	\newcommand{\nbc}[3]{}
 	
 }
\newcommand\mg[1]{\nbc{MG}{#1}{purple}}

\newcommand{\ie}{\emph{i.e.},\xspace}
\newcommand{\eg}{\emph{e.g.},\xspace}

\AtBeginDocument{\hypersetup{citecolor=DodgerBlue4}}

\usepackage[all]{nowidow}
\usepackage{mdframed}
\usepackage{enumitem}
\definecolor{light-gray}{gray}{0.97}  
\newcounter{rmd}
\newenvironment{rmd}[1][]{
    \refstepcounter{rmd}
    \mdfsetup{innertopmargin=0.7Em,
              linecolor=gray!20,
              linewidth=2pt,
              topline=true, 
              backgroundcolor=light-gray,
              nobreak=true}
    \begin{mdframed}[] {} \relax
    }{\end{mdframed}}

\usepackage{collcell}
\usepackage{xfp}

\newcommand{\ApplyGradientK}[1]{%
  \ifnum#1>10 \colorbox{blue!70!white}{\makebox[1.2em][l]{\textcolor{white}{#1}}}%
  \else \ifnum#1>5 \colorbox{blue!30!white}{\makebox[1.2em][c]{#1}}%
  \else \ifnum#1>0 \colorbox{blue!10!white}{\makebox[1.2em][c]{#1}}%
  \else \colorbox{blue!0!white}{\makebox[1.2em][c]{#1}}%
  \fi\fi\fi
}

\newcommand{\ApplyGradientL}[1]{%
  \ifnum#1>10 \colorbox{red!70!white}{\makebox[1.2em][l]{\textcolor{white}{#1}}}%
  \else \ifnum#1>5 \colorbox{red!30!white}{\makebox[1.2em][c]{#1}}%
  \else \ifnum#1>0 \colorbox{red!10!white}{\makebox[1.2em][c]{#1}}%
  \else \colorbox{red!0!white}{\makebox[1.2em][c]{#1}}%
  \fi\fi\fi
}

\newcommand{\ApplyGradientM}[1]{%
  \ifnum#1>10 \colorbox{green!70!white}{\makebox[1.2em][l]{\textcolor{white}{#1}}}%
  \else \ifnum#1>5 \colorbox{green!30!white}{\makebox[1.2em][c]{#1}}%
  \else \ifnum#1>0 \colorbox{green!10!white}{\makebox[1.2em][c]{#1}}%
  \else \colorbox{green!0!white}{\makebox[1.2em][c]{#1}}%
  \fi\fi\fi
}

\newcommand{\ApplyGradientN}[1]{%
  \ifnum#1>10 \colorbox{orange!70!white}{\makebox[1.2em][l]{\textcolor{white}{#1}}}%
  \else \ifnum#1>5 \colorbox{orange!30!white}{\makebox[1.2em][c]{#1}}%
  \else \ifnum#1>0 \colorbox{orange!10!white}{\makebox[1.2em][c]{#1}}%
  \else \colorbox{orange!0!white}{\makebox[1.2em][c]{#1}}%
  \fi\fi\fi
}

\newcommand{\ApplyGradientX}[1]{%
  \ifnum#1>10 \colorbox{blue!70!white}{\parbox{1em}{\hfill\textcolor{white}{#1}}}%
  \else \ifnum#1>5 \colorbox{blue!30!white}{\parbox{1em}{\hfill#1}}%
  \else \ifnum#1>0 \colorbox{blue!10!white}{\parbox{1em}{\hfill#1}}%
  \else \colorbox{blue!0!white}{\parbox{1em}{\hfill#1}}%
  \fi\fi\fi
}

\newcommand{\ApplyGradientY}[1]{%
  \ifnum#1>10 \colorbox{green!70!white}{\parbox{1em}{\hfill\textcolor{white}{#1}}}%
  \else \ifnum#1>5 \colorbox{green!30!white}{\parbox{1em}{\hfill#1}}%
  \else \ifnum#1>0 \colorbox{green!10!white}{\parbox{1em}{\hfill#1}}%
  \else \colorbox{green!0!white}{\parbox{1em}{\hfill#1}}%
  \fi\fi\fi
}

\newcommand{\ApplyGradientZ}[1]{%
  \ifdim \fpeval{#1 > 100.0} pt > 0pt
    \colorbox{orange!70!white}{\makebox[1.5em][c]{\textcolor{black}{#1}}}%
  \else\ifdim \fpeval{#1 > 50.0} pt > 0pt
    \colorbox{orange!30!white}{\makebox[1.5em][c]{#1}}%
  \else\ifdim \fpeval{#1 > 10.0} pt > 0pt
    \colorbox{orange!10!white}{\makebox[1.5em][c]{#1}}%
  \else
    \colorbox{orange!0!white}{\makebox[1.5em][c]{#1}}%
  \fi\fi\fi
}

\newcommand{\ApplyWhiteBackground}[1]{%
  \ifnum#1>10 \colorbox{white!70!white}{\makebox[1.4em][c]{\textcolor{black}{#1}}}%
  \else\ifnum#1>5 \colorbox{white!30!white}{\makebox[1.4em][c]{#1}}%
  \else\ifnum#1>0 \colorbox{white!10!white}{\makebox[1.4em][c]{#1}}%
  \else \colorbox{white!0!white}{\makebox[1.4em][c]{#1}}%
  \fi\fi\fi
}

\newcommand{\ApplyWhiteBackgroundI}[1]{%
  \ifnum#1>10 \colorbox{white!70!white}{\makebox[1.4em][c]{\textcolor{black}{#1}}}%
  \else\ifnum#1>5 \colorbox{white!30!white}{\makebox[1.4em][c]{#1}}%
  \else\ifnum#1>0 \colorbox{white!10!white}{\makebox[1.4em][c]{#1}}%
  \else \colorbox{white!0!white}{\makebox[1.4em][c]{#1}}%
  \fi\fi\fi
}

\newcolumntype{T}{>{\collectcell\ApplyWhiteBackground}r<{\endcollectcell}}
\newcolumntype{I}{>{\collectcell\ApplyWhiteBackgroundI}r<{\endcollectcell}}

\newcolumntype{K}{>{\collectcell\ApplyGradientK}r<{\endcollectcell}}
\newcolumntype{L}{>{\collectcell\ApplyGradientL}r<{\endcollectcell}}
\newcolumntype{M}{>{\collectcell\ApplyGradientM}r<{\endcollectcell}}
\newcolumntype{N}{>{\collectcell\ApplyGradientN}r<{\endcollectcell}}
\newcolumntype{R}[1]{>{\raggedleft\arraybackslash}p{#1}}

\newcolumntype{X}{>{\collectcell\ApplyGradientX}r<{\endcollectcell}}
\newcolumntype{Y}{>{\collectcell\ApplyGradientY}r<{\endcollectcell}}
\newcolumntype{Z}{>{\collectcell\ApplyGradientZ}r<{\endcollectcell}}

\journal{Computers \& Security}

\title{Metaverse Security and Privacy Research: A Systematic Review}

\begin{document}

\begin{frontmatter}




\author[label1]{Argianto Rahartomo}
\ead{argianto.rahartomo@tu-clausthal.de}
\author[label2]{Leonel Merino}
\ead{leonel.merino@uc.cl}
\author[label1]{Mohammad Ghafari}
\ead{mohammad.ghafari@tu-clausthal.de}

\affiliation[label1]{organization={Technische Universität Clausthal}, 
            city={Goslar},
            country={Germany}}

\affiliation[label2]{organization={School of Design, School of Engineering, Pontificia Universidad Católica de Chile}, 
            city={Santiago},
            country={Chile}}


\begin{abstract}
The rapid growth of metaverse technologies, including virtual worlds, augmented reality, and lifelogging, has accelerated their adoption across diverse domains. This rise exposes users to significant new security and privacy challenges due to sociotechnical complexity, pervasive connectivity, and extensive user data collection in immersive environments. We present a systematic review of the literature published between 2013 and 2024, offering a comprehensive analysis of how the research community has addressed metaverse-related security and privacy issues over the past decade. We organize the studies by method, examined the security and privacy properties, immersive components, and evaluation strategies. Our investigation reveals a sharp increase in research activity in the last five years, a strong focus on practical and user-centered approaches, and a predominant use of benchmarking, human experimentation, and qualitative methods. Authentication and unobservability are the most frequently studied properties. However, critical gaps remain in areas such as policy compliance, accessibility, interoperability, and back-end infrastructure security. We emphasize the intertwined technical complexity and human factors of the metaverse and call for integrated, interdisciplinary approaches to securing inclusive and trustworthy immersive environments.
\end{abstract}


\end{frontmatter}



\section{Introduction}
\label{sec:introduction}
The metaverse offers unprecedented opportunities to reshape how we engage with digital and physical spaces, powered by advancements in augmented reality, virtual reality, and artificial intelligence. However, as this ecosystem evolves, so do the risks associated with security and privacy vulnerabilities~\citep{huang2023, rahartomo2025dilemma}. Addressing these challenges is imperative, especially as the metaverse begins to permeate critical domains such as education, healthcare, and commerce. 

We conducted a systematic study of 114 metaverse security and privacy papers published between 2013 and 2024, aiming to 
investigate research methods in this domain.
We explored the types of studies conducted, the properties emphasized, the research strategies adopted, and the scope of their evaluations.
To ensure the credibility of our findings, we also analyzed common threats to validity across the reviewed studies, identifying recurring risks that may affect the reliability, generalizability, or interpretation of results in metaverse security and privacy research.

We found that the number of publications
continues to grow, with a noticeable shift toward technique-driven and evaluation-focused studies.
Authentication and confidentiality 
(\ie 22 and
12 of 68 articles), are
the most widely explored security properties
, underscoring their critical role in safeguarding
the metaverse.
Privacy aspects such as unobservability (hiding a user's actions) and content awareness (ensuring appropriate access to virtual environments) also garnered significant attention 
(\ie 14 and 12 of 46 articles)
, highlighting a growing concern for user privacy in immersive settings.
A substantial proportion of studies (71\%) employed multiple research strategies to address the unique complexities of immersive environments
, such as evaluating real-world usability, capturing subjective user experiences, and measuring system performance.
The most commonly used strategies were benchmarking (30\%), human experimentation (26\%), and interviews (14\%).
In total, 69\% of the metaverse studies involved human participants, with a median sample size of 25 participants. These studies typically evaluated user performance through metrics such as task completion times and task accuracy.
Our analysis of threats to validity across the reviewed studies uncovered several recurring concerns: limited generalizability stemming from small or unrepresentative samples, insufficient methodological transparency that undermines the reliability of findings, and disalignment between theoretical constructs and their practical implementation.

Our
investigation
also highlighted several critical gaps in metaverse security and privacy research
areas that warrant further investigation.

\textit{Infrastructure and Network Protocols:}
While much of the existing research focuses on 
virtual and augmented reality (
AR / VR
) 
hardware, security\del{,} and privacy concerns related to back-end infrastructures and network communication protocols\del{,} have been significantly underexplored, posing potential vulnerabilities in large-scale metaverse systems.
Infrastructure and network protocols are essential to the functioning of metaverse systems, but have not received much attention in current research. Some challenges, such as inadequate encryption for real-time communication and inconsistent identity management, pose risks to user privacy and system security. Although emerging solutions, such as more secure communication standards and the implementation of artificial intelligence that preserves privacy, show promise, these approaches are still in the early stages and need further investigation.

\textit{Interoperability:}
Only three studies addressed interoperability between different metaverse platforms. This leaves a critical research gap around how data is exchanged and managed securely across diverse virtual environments, with potential implications for user privacy.
In fact, interoperability in the metaverse is based not just on asset portability but also on deep integration at the back-end infrastructure and network protocol level. Current metaverse platforms, such as Horizon Worlds, Decentraland, and Roblox, rely on incompatible server architectures and data models, making real-time cross-platform interaction difficult.
Without a
standard
protocol stack or federated identity system, even basic cross-platform movement could compromise user privacy or system integrity.

\textit{Accessibility for Disabilities:}
The accessibility of the metaverse for users with disabilities or disorders is 
an overlooked topic.
Despite the promise of immersive technologies to enable new forms of interaction, participation, and presence,
only one
study
~\citep{metaversepaper089} explicitly addressed the needs of individuals with low vision. This limited attention is particularly concerning given the potential of the metaverse to bridge or widen digital divides
, and the unique security risks this user group faces.
In addition, immersive devices often lack accessibility features by default, and existing privacy controls may not be adaptable to different perceptual or interaction needs.

\textit{Regulatory and Compliance Concerns:}
Many existing studies overlook critical aspects of metaverse governance, including regulatory frameworks and compliance measures that are essential for protecting user privacy. This oversight risks creating significant gaps in the legal and ethical protections available to users. The metaverse presents substantial social and ethical challenges, especially with respect to data privacy, identity, and digital autonomy. Immersive and persistent virtual environments collect extensive personal and behavioral data, prompting serious concerns about surveillance, informed consent, and data control. The flexible nature of digital identity, where avatars can differ significantly from real-world user personas, further complicates issues of accountability, authenticity, and representation. In the absence of well-defined regulations and ethical standards, these concerns may erode user trust and hinder the widespread adoption of metaverse technologies.

\textit{Scalability and Performance Under Load:}
As the metaverse scales in both user base and technological complexity, ensuring consistent performance becomes a critical challenge. High concurrency, real-time interactions, and the integration of diverse media, such as rich 3D environments and data streams, place immense pressure on the network infrastructure, rendering systems vulnerable to latency, lag, and instability. Additionally, managing distributed computation and maintaining synchronization across heterogeneous devices further complicates scalability.

In summary, 
our
study offers a comprehensive analysis of current trends in metaverse security and privacy research, identifies significant gaps, and highlights critical issues that warrant further investigation. Our findings set the stage for future research, informing efforts to strengthen the security and privacy foundations of this digital ecosystem. 

We call on researchers in the field to prioritize accessibility by exploring inclusive design principles, evaluating assistive technologies in extended reality (XR) settings, and developing security and privacy models that accommodate diverse abilities, ensuring equitable and trustworthy participation in metaverse environments.
We also urge the community to address regulatory and compliance challenges by examining how different user groups experience risks and by guiding the development of strong privacy protections, identity management systems, and inclusive governance frameworks.
Finally, we encourage research on scalability and performance, particularly through adaptive load balancing and decentralized architectures, such as blockchain or peer-to-peer systems, which offer promising solutions to meet the growing demands of metaverse platforms.

The remainder of this 
article
is organized as follows.
We describe our research methodology in Section~\ref{methodology}.
We present and discuss our findings in Section~\ref{results}.
We
explain
threats to the validity of this study in Section~\ref{threats} and conclude the article in Section~\ref{conclusion}.

\section{Methodology}
\label{methodology}

\begin{figure}[ht]
    \centering
    \includegraphics[width=\linewidth]{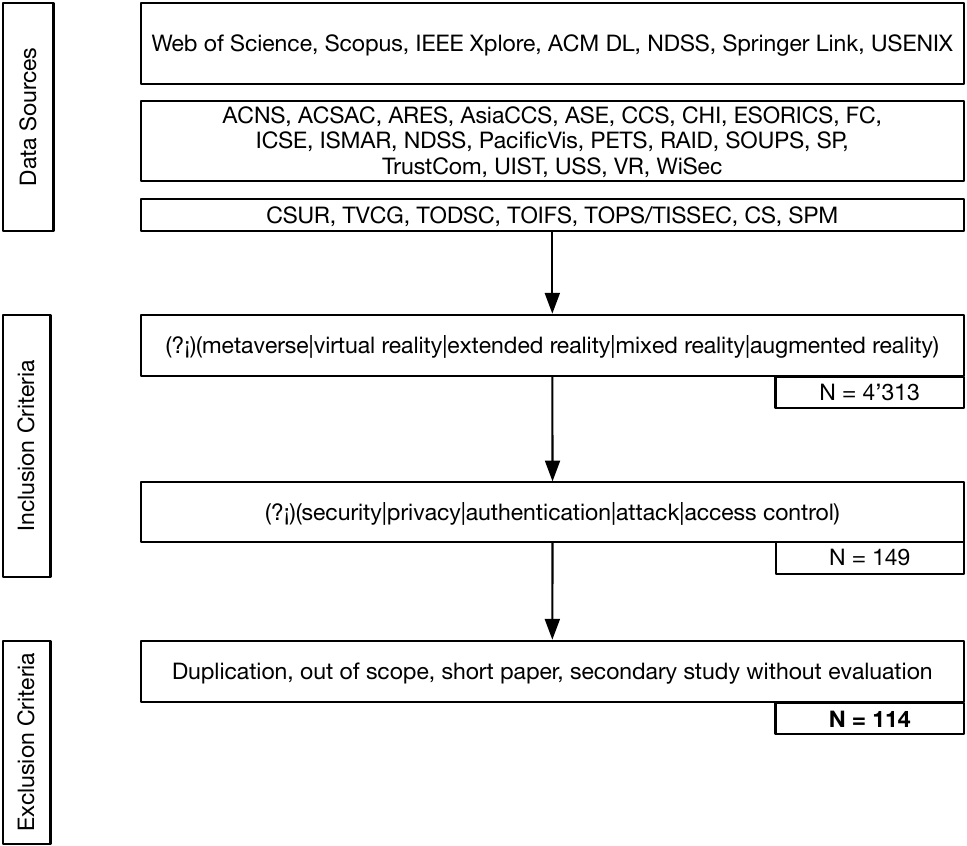}
    \caption{The process for selecting 114 papers that evaluate immersive technologies in security and privacy.}
    \label{fig:methodologyprocessdiagram}
\end{figure}

We followed the systematic review of the literature (SLR) method, which is highly popular in Software Engineering~\citep{kitchenham_2007}. The diagram shown in Figure~\ref{fig:methodologyprocessdiagram} represents the process we followed for data selection (\ie identify data sources and establish inclusion and exclusion criteria). Next, we discuss the data extraction process. 

\subsection{Data selection}


We seek articles that describe evaluations involving immersive technologies (\eg metaverse, virtual reality, augmented reality) within the domain of security and privacy. To ensure the high quality of primary studies, we examined the best publications, including conference and journal proceedings, on security and privacy, software engineering, and human-computer interaction. 

\emph{Data sources.} We curated a list of venues A *, A and B based on the Core Ranking for conferences\footnote{\url{http://portal.core.edu.au/conf-ranks/}} and journals\footnote{\url{http://portal.core.edu.au/jnl-ranks/}}. 

\emph{Inclusion criteria.} We used popular digital libraries and search engines (\ie Web of Science\footnote{\url{https://mjl.clarivate.com/home}}, Scopus\footnote{\url{https://www.scopus.com/search/form.uri}}, IEEE Xplore\footnote{\url{https://ieeexplore.ieee.org/xplore/home.jsp}}, ACM DL\footnote{\url{https://dl.acm.org/}}, Springer Link\footnote{\url{https://link.springer.com/}}) to identify articles suitable for the scope of the study. Articles published in \href{https://www.usenix.org/}{USENIX} and \href{https://www.ndss-symposium.org/}{NDSS} were not included in these databases, but are made accessible on their respective websites. 
We selected articles published between 2013 and 2024. 
We first included papers that contain (in titles, abstracts, or keywords) a metaverse-related keyword, and then included papers that also contain a keyword related to the security and privacy domain.
We focused our selection on the top-ranked venues in the security and privacy fields, including USS, SP, CCS, NDSS, CRYPTO, and JoC. From this analysis, after excluding common stop words and nonspecific terms such as \emph{user}, \emph{data}, \emph{system}, and \emph{device}, we identified the five most recurring words relevant to security: \emph{security}, \emph{privacy}, \emph{authentication}, \emph{attack}, and \emph{access control}.
We used them to narrow down the results and focus on papers more relevant to the security and privacy domain. 
We developed a Python script\footnote{
\url{https://doi.org/10.5281/zenodo.15738685}
}
to automate metadata extraction. We used the Computer Science Bibliography (DBLP)\footnote{\url{https://dblp.org/db/}} to obtain the title of articles when direct access to metadata was not available. Next, we used the titles to retrieve the metadata of the articles using Google Scholar\footnote{\url{https://scholar.google.com/}}. Then, we manually verified the results to ensure the completeness of the metadata.

\emph{Exclusion criteria.} Next, we excluded duplicate results (\ie articles returned by more than one source), short articles for which we included an extended version, and secondary studies without evaluations. We also excluded a study that served primarily as a call for submissions of papers to a conference and some studies that present a systematic taxonomy and classification of specific topics without discussing any form of evaluation method. 
Finally, we selected 114 papers.
\vspace{0cm}
\subsection{Data Extraction}
For each article, we categorize \ins{the details of the metaverse considered,} the type of study, the security and privacy \chg{concerns, the research strategy used, the metaverse components considered}{properties}, \ins{the research strategy used, the characteristics of the evaluations carried out}, \chg{the type of reality adopted to alleviate these concerns and the characteristics of the evaluations carried out}{ and threats to validity}.
The first two authors of the article participated in the extraction process. We calibrated our assessments with a small subset of papers. We discussed the results and solved conflicting classifications. Finally, we independently analyzed the 114 included articles, compared the results, and agreed on the classifications by consensus.

\subsubsection{Metaverse Focus}
To define the focus of the metaverse, we categorize the components outlined in research and the various forms of reality.

\paragraph{Component} Since there is no\ins{t} 'one' metaverse, we used a popular taxonomy~\citep{smart2007cross} to classify the metaverse components described in studies\ins{. We classified the components} into one of the following four categories.
Notice that the \emph{augmented reality} component is not restricted to a specific type of reality, as we will discuss next.
\begin{itemize}
    \item{Virtual World\ins{s}} are digital environments in which users interact with each other and with the surroundings through avatars. An example is a study of Lin about preserving avatar's authenticity~\citep{metaversepaper097}.
    \item{Lifelogging} involves the ongoing process of recording one's own \ins{data of} experiences and daily activities, often using wearable devices. One example is a study \chg{of}{by} DeVrio about utilizing the smartwatch for real-time tracking of user's gesture and activity~\citep{metaversepaper087}.
    \item{Augmented Reality} refers to the process of superimposing digital information or objects on the real world\del{ using devices}. An example is a study \chg{of}{by} Lehman about the privacy risks of augmented reality~\citep{metaversepaper039}.
    \item{Mirror World} is about creating digital replicas of the real world, accurately reproducing real-life locations, events, and objects. An example is a study \chg{of}{by} Maddali and Lazar on how to understand the social context to create meaningful reconstructions of physical spaces for remote instruction and interaction~\citep{metaversepaper098}.
\end{itemize}

\paragraph{Reality Type}
We adopted the definitions of the types of reality described in the mixed reality continuum~\citep{milgram1994taxonomy}. We extend\ins{ed} the definitions to include extended reality~\citep{esen_metaverse_2023}
We classified each article on the basis of explicit descriptions of various types of reality. Those articles mentioning the term \emph{metaverse} without specifying a particular type of reality were classified as \emph{XR}~\citep{metaversepaper024}.

\begin{itemize}
    \item{Virtual reality} (VR) is an artificial digital environment that provides an immersive experience to users, often simulating real-world interactions through multisensory feedback. Examples of VR include interactive 3D games like Minecraft VR and professional applications such as virtual training simulators for education and healthcare.
    %
    \item{Augmented Reality} (AR) is a technology that enhances the real world by overlaying digital information such as sound, video, graphics, or GPS data on it, and an example of AR is the Timetraveler app, which allows users to experience historical events through augmented visuals and sounds at specific locations.
    \item{Mixed Reality} (MR) is an interactive technology that blends real-world and virtual elements, where digital and physical objects coexist and interact in real-time, and an example of MR is the Microsoft HoloLens which offers a hybrid of real and virtual experiences by projecting holograms into the user's environment.
    \item{Extended Reality} (XR) is a collective term for immersive technologies such as Virtual Reality (VR), Augmented Reality (AR) and Mixed Reality (MR) that enhance or replace the physical world with digital elements, exemplified by headsets that overlay holograms on the real world or fully immersive simulations.
\end{itemize}

\subsubsection{Study Type}
We classified the 114 research articles based on Munzner's framework~\citep{Munzner2008}, which we adjusted to the scope of our study, into one of five types:

\paragraph{Applications} Studies that describe the use of existing metaverse techniques in order to solve a concrete and relevant problem in the security and privacy domain. That is, these papers do not present novel techniques but focus on design decisions during the development process. Artifacts are often available to promote the reproducibility of the results. When found, we classified these artifacts into the following categories:
\begin{itemize}
    \item{Application Code (AC)} refers to the source code and accompanying documentation necessary for the installation and execution of a software program;
    \item{Experimental Data (ED)} refers to supplemental materials, including datasets and other artifacts that are utilized in experiments; and
    \item{Executable Applications (EX)} are compiled programs that can be run but cannot be altered.
\end{itemize}

\paragraph{Evaluations} Studies that concentrate on the assessment of applications, often involving users. Evaluations can involve various methodologies, such as case studies, user studies, and controlled experiments. Depending on the goal of an evaluation, they can collect and analyze data such as user performance or user experience.

\paragraph{Models} Papers that, for example, introduce formalisms to describe new abstractions, definitions, or terminology to characterize methods or analyze phenomena. Models also include taxonomies for classifications to understand a particular subject and can involve commentaries in which the authors of a paper present arguments to support a position.

\paragraph{Systems} Studies that focus on the architecture of a framework or toolkit designed to support the development of applications. Unlike applications, systems do not address a particular security and privacy property but rather a type of problem.

\paragraph{Techniques} Papers that provide comprehensive explanations of the supporting algorithms and technical descriptions of the implementations. Often, techniques are aimed at efficiency and performance (\eg to reduce memory usage, reduce processing time, or improve overall performance) and are evaluated using benchmarks to compare with state-of-the-art techniques.
\vspace{1em}

\subsubsection{Security and Privacy Property}
\label{subsec:secprivproperties}
To classify the articles, we used the security and privacy properties that arise when building software, proposed in a previous popular study~\citep{deguzman2019mr}. 
We determined the most prominent category by analyzing the descriptions of the threat or privacy model, which we often found explicitly in a dedicated section of articles.
    
\paragraph{Authentication} It checks the legitimacy of users who access the metaverse device or service, allowing only authenticated users to progress to the identification and authorization steps. An exemplary study \chg{of}{by} George investigates the use of 3D spatial user behavior as an authentication method for smart homes in virtual and real realities~\citep{metaversepaper078}.

\paragraph{Confidentiality} Protects sensitive data from unauthorized access by implementing rigorous access controls to personal and identifiable information. A notable study highlights the assumed confidentiality in VR interactions and advocates for enhanced security protocols~\citep{metaversepaper020}.
  
\paragraph{Authorization and access control} It requires that actions and processes be initiated only by verified parties with appropriate access levels, guaranteeing that only authorized apps can interact with specified data or objects. Current authorization models for mixed reality platforms expose vulnerabilities in 3D spatial maps, which shows the need for stronger access control mechanisms~\citep{metaversepaper019}.

\paragraph{Integrity} It relates to ensuring that data and processes in metaverse environment remain unchanged and accurate, allowing for the proper detection and display of virtual augmentations without unauthorized changes. For example, the integrity of avatars can be enhanced through visual indicators to identify abusive ad service providers in virtual environments~\citep{metaversepaper097}.

\paragraph{Identification} It entails assigning each action within metaverse system to a specific entity, hence easing access management and avoiding unwanted actions by ensuring all participants are recognized. For example, the characteristics of the user's gait can be used for their identification, which requires a balance between recognition accuracy and preservation of privacy in virtual environments~\citep{metaversepaper041}.

\paragraph{Availability} It emphasizes the importance of ongoing access to data and services within a metaverse application, to prevent attackers from impeding these resources. An exemplary study of Rovira proposed a high availability system architecture to guarantee consistent content access~\citep{metaversepaper111}.

\paragraph{Non-repudiation} It ensures that if an action is taken or data are modified within metaverse application, the entity responsible cannot deny their involvement, hence establishing accountability via digital evidence. For example, data provenance can be used to trace and display cyberattack patterns, providing detailed forensic trails from initial reconnaissance to system exploitation, thus ensuring non-repudiation~\citep{metaversepaper052}.

\paragraph{Unobservability \& Undetectability} Protects entities' presence or activity from detection by attackers, ensuring that activities remain secret and indistinguishable from noise. For example, personal data can be concealed by playing in an 'escape room' scenario, demonstrating the effectiveness of unobservability and undetectability in protecting privacy~\citep{metaversepaper042}.

\paragraph{Content Awareness} It ensures that users are completely aware of the nature and sensitivity of the data they share, fostering transparency and informed consent. In fact, there is a need to integrate user perspectives and privacy considerations in XR design, in particular, how content awareness can enhance user engagement and data protection~\citep{metaversepaper098}.

\paragraph{Anonymity \& Pseudonymity} It enables entities to separate or disguise their identities from data or actions, providing privacy while preventing adversaries from tracing activities back to individuals. For example, the GaitLock system integrates innovative gait recognition techniques based on dynamic time warping and sparse representation classifier to advance pseudonymity~\citep{metaversepaper032}.

\paragraph{Policy \& Consent Compliance} \chg{It e}{E}nsures a system to comply with specified privacy and security rules, ensuring that user rights are respected and enforced. For example, there are significant privacy topics Oculus with VR applications, which requires a thorough re-assessment of policy frameworks to better protect user data~\citep{metaversepaper012}.

\paragraph{Unlinkability} Prevents adversaries from linking an entity to specific data or behaviors, protecting privacy by hiding linkages between user activities. An exemplary study \chg{of}{by} Patan and \del{M. }Parizi, combines encryption with blockchain technology to protect against data breaches and obscures personal data from unauthorized inference~\citep{metaversepaper044}.
%

\paragraph{Plausible Deniability} It allows entities to plausibly deny involvement in actions or data storage while providing privacy protection against data origin tracing. For example, spatial augmented reality (SAR)~\citep{metaversepaper072}, proposes a new methodological strategy to achieve plausible deniability by offering robust privacy protections in digital environments.

\subsubsection{Research Strategy} 
We identify the research strategies adopted in studies based on the empirical software engineering standards of the ACM~\citep{acmstandards1, acmstandards2, acmstandards3}.

\paragraph{Benchmarking} Comparison of the efficacy of various methods, tools, or techniques in real-world contexts. Typically, it compares the study with previous works. For example, benchmarks can be used to evaluate the reliability of an authentication model over time~\citep{metaversepaper009}.

\paragraph{Human Experimentation} Engages users in testing or model development and observes the effects of deliberate interventions under controlled conditions to study aspects of reality. For example, an experiment can help to assess the experiences of adolescents and potential security threats from various perspectives~\citep{metaversepaper050}.

\paragraph{Qualitative Survey (Interviews)} Consists of semi-structured or open-ended interviews for data collection. It is explicitly mentioned in the papers. For example, interviews could be used to investigate user reactions to perceptual manipulation attacks~\citep{metaversepaper018}.

\paragraph{Questionnaire Surveys} Collects responses to a structured series of questions using digital or on-line questionnaires. It is also explicitly mentioned in the papers. For example, questionnaires can be used to explore the usability and security of authentication mechanisms in a virtual reality (VR) setting~\citep{metaversepaper066}.

\paragraph{Data Science} Apply data-centered methodologies, including machine learning algorithms and models for data analysis and interpretation, to examine software engineering phenomena. For example, data science methodologies such as machine learning and deep learning \chg{were}{have been} instrumental in analyzing motion data from more than 50,000 users~\citep{metaversepaper022}.

\paragraph{Engineering Research} Focuses on the creation and evaluation of technological artifacts, including algorithms, systems, tools, and other computer-based technologies. For example, this approach was used to examine the APIs of wallets in different applications and websites to identify possible data breaches~\citep{metaversepaper024}.

\paragraph{Meta Science}Analyzes research methodologies or offers guidelines for the execution of research, including taxonomy studies. For example, a taxonomy was created to categorize authentication techniques for AR headsets~\citep{metaversepaper045}.


\subsubsection{Evaluation Scope}

We review the scope of evaluations in metaverse security and privacy by collecting data from scenarios in which evaluations are conducted, extracting their quality focus, characteristics of the subjects involved, and the use of research artifacts.

\paragraph{Scenario}
To characterize evaluations, we classify the scenarios involved into four types\del{ (\ie algorithm performance, user experience, understanding environments and work practices, and user performance)}.

\begin{itemize}
\item{Algorithm Performance} entails employing quantitative methods, frequently through benchmarks, to assess the effectiveness and occasionally the efficiency of algorithms, focusing on aspects such as speed and resource usage. For example, they compare results with other approaches and evaluate constraints and behaviors across different data volumes and complexities. Often, such evaluations analyze the algorithm's relative speed, scalability, and performance in extreme circumstances.
\item{User Experience} focuses on capturing the internal state of users when interacting with a technology. \chg{They can}{This type of scenarios} collect data \chg{that range}{ranging} from initial impressions to long-term usage \chg{thoughts}{assessments}. Such evaluations usually involve the use of surveys and interviews, which often collect data using metrics such as the Likert scale. The data collected can involve \chg{the}{aspects such as} usability, intuitiveness, trust, and overall satisfaction of the tool, as well as the identification of potential gaps in the tool's functionality and design. Such evaluations might range from informal feedback meetings to systematic usability testing and thorough field observations, providing immediate and in-depth insights on user experiences.
\item{Understanding Environments and Work Practices} relate to assessments to obtain requirements with the goal of understanding the security and privacy needs of users and organizations before developing a metaverse-related approach. Common data collection methods used to understand environments and work practices are \ins{observations,} surveys\ins{,} and interviews.
\item{User Performance} assesses measurable factors such as completion time and error rates, alongside quantifiable qualitative assessments. These evaluations may involve user studies that transform real-world tasks into constrained activities, with the aim of either widespread participation to generalize the findings or concentrating on smaller cohorts to deeply understand a specific phenomenon. 
\end{itemize}

\paragraph{Quality Focus} It refers to the key aspects of primary concern in an experiment.
These include the specific \chg{effects or outcomes that you want to understand or}{variables that will be} measure\ins{d}, such as the accuracy of the results, the usability of the system, the robustness of its adaptability, and the speed \chg{at}{with} which tasks can be completed. 
For example, the quality focus of the study by Liebers is \ins{on} correctness and robustness~\citep{metaversepaper065}.

\paragraph{Subject}
We collect data from the subject being evaluated, which frequently involves study participants. For example, the study by Lin evaluates the authenticity of the avatar in the context of 60 participants~\citep{metaversepaper097}. \chg{However}{In addition}, in evaluations that do not involve participants, we collected information on software systems and data sets. 


\paragraph{Artifact}
We collect involved artifacts such as compiled applications and their source code, as well as experimentation data sets. In addition, we collect information about frameworks and programming languages described in the evaluations. Usually, we find these data in the implementation sections \chg{and}{or by} analyzing project repositories. In \del{the} repositories\ins{,} we found information such as programming languages and frameworks. We \chg{hope that}{consider} this information \del{will }help\ins{ful for} practitioners and researchers in the field \ins{to} identify actionable tools and assess their maturity level.

\subsubsection{Threat to Validity}
In addition, we collect data on threats that often affect the validity of the results of security and privacy studies in the metaverse. Such threats represent risks that can result in findings that are not accurate or trustworthy.
We differentiate \del{the }threats to validity (TTV) from \del{the }limitations, as limitations refer to constraints or shortcomings in the design or execution of the study. Limitations could affect the interpretation or applicability of the results, but do not necessarily invalidate them. We concentrate on threats to validity because they affect the accuracy or credibility of findings, while we exclude limitations because they affect the extent or conditions under which the findings could be useful. 


We note that TTVs are occasionally stated explicitly, frequently in a specific section\chg{, w}{. W}hile in other studies there may be implicit information that allows us to deduce TTVs. We plan to label explicit and implicit data. 
To ensure validity, we involved five people from our institution who double-checked the classification of 10\% of the collected studies.
%
%
For each study, we classify TTVs into four categories based on popular frameworks~\citep{wohlin2024experimentation,campbell1979}. Specifically, we used the following categories: i) threats to internal validity, ii) threats to external validity, iii) threats to conclusion validity, and iv) threats to construct validity.
\vspace{0em}
\paragraph{Threats to internal validity} 
Threats to internal validity are factors that can lead to incorrect conclusions about causal relationships in a study, such as whether changes in the independent variable truly caused the observed changes in the dependent variable. Such threats arise when alternative explanations for the results are possible due to flaws in the study design or execution. Frequent examples involve participants who \del{either }show improvement because of practice or experience adverse effects (such as fatigue or boredom) during an experiment\chg{,}{. Other examples, include} unforeseen events or uncommon circumstances arising during the study, or challenges related to the selection of participants and dropout rates.
%

\paragraph{Threats to external validity} 
Threats to external validity are factors that limit the generalizability of study findings beyond the specific conditions of the investigation. That is, they pose a risk, as the outcomes might not be valid in various contexts, populations, or time periods. Common examples involve participants who are not representative of the larger population, the setting of the study (\eg lab, online) may not reflect real-world conditions, or the findings may only be valid at the time the study was conducted.


\paragraph{Threats to conclusion validity} 
Threats to the validity of conclusions are factors that could affect the precision and reliability of conclusions about the relationship between variables, especially whether a relationship exists at all. That is, these threats influence the correct interpretation of statistical evidence regarding a relationship.
%
Common threats arise in studies that lack sufficient sample size to detect a true effect, statistical tests that assume certain conditions (\eg normality, independence), which may not be valid, or when treatments are applied differently among participants.

\paragraph{Threats to construct validity} 
Threats to construct validity are factors that diminish the extent to which a research study precisely measures or manipulates the theoretical ideas (constructs) it claims to examine. That is, even if an effect is observed in a study, these threats challenge whether the effect pertains to the actual concept \del{intended to be }investigated. Common examples occur when the concept studied is either vaguely or inconsistently defined, measured, or manipulated using a single method, or evaluated with just one measurement technique.

\section{Results and Discussion}
\label{results}
\begin{figure}[!h]
   \centering
   \includegraphics[width=\linewidth]{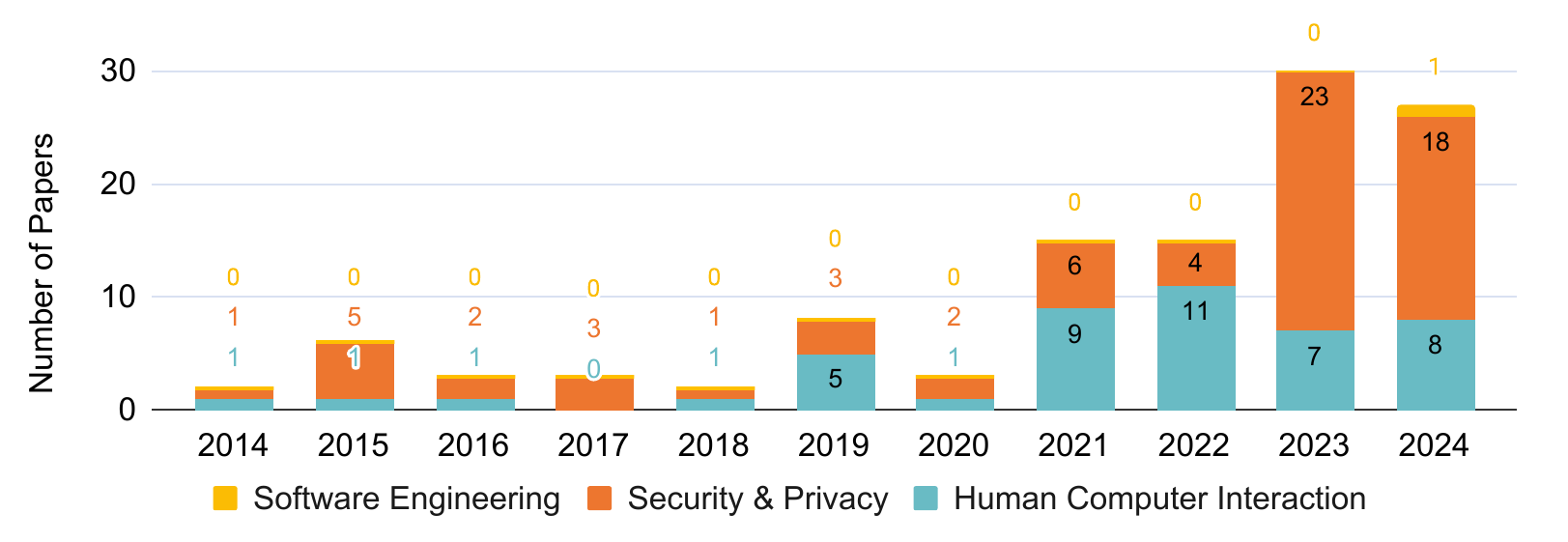}
   \caption{The 114 included papers by publication year and domain.}
   \label{fig:NumberOfPapersByYear}
\end{figure}

Table~\ref{tab:metaversepaper1} lists the 114 selected articles, and Figure~\ref{fig:NumberOfPapersByYear} shows their distribution over time. 
In the stacked bar chart, the articles published in security \& privacy (SP) venues are colored orange, software engineering (SE) venues are colored yellow, while those in human-computer interaction venues (HCI) appear in light blue. 
We observe that most papers have been published in the last five years (\ie 78.95\%).
In the past five years, there has been a significant increase in research focused on metaverse security and privacy. 
This improvement can be driven by metaverse unique risks, such as user authentication challenges~\citep{metaversepaper006} and the potential for user impersonation~\citep{metaversepaper037}. 
In addition, emerging threats such as harassment~\citep{dwivedi_exploring_2023}, identity theft, and data misuse~\citep{pooyandeh2022} have also contributed to increased attention. 
In HCI venues, the number of papers has increased moderately in the last three years. 
The growth in HCI research is supported by advances in hardware, including the release of Meta Quest 2 in 2020, as well as a growing engagement in the community of VR/AR software developers.
\begin{table*}[ht!]
\centering
\tiny
\renewcommand\arraystretch{0.1}
    \caption{List of 114 Included Papers.}
\label{tab:metaversepaper1}
\begin{tabular}{llll}
    \textbf{Year} & &\textbf{Title}&\textbf{Venue} \\ \hline

2024&\citep{metaversepaper113}&Understanding Parents' Perceptions and Practices Toward Children's Security and Privacy in Virtual Reality&SP \\
 &\citep{metaversepaper114}&Eavesdropping on Controller Acoustic Emanation for Keystroke Inference Attack in Virtual Reality&NDSS \\
 &\citep{metaversepaper115}&That Doesn’t Go There: Attacks on Shared State in Multi-User Augmented Reality Applications&USS \\
 &\citep{metaversepaper116}&Penetration Vision through Virtual Reality Headsets: Identifying 360-degree Videos from Head Movements&USS \\
 &\citep{metaversepaper117}&Can Virtual Reality Protect Users from Keystroke Inference Attacks?&USS \\
 &\citep{metaversepaper118}&When the User Is Inside the User Interface: An Empirical Study of UI Security Properties in Augmented Reality&USS \\
 &\citep{metaversepaper119}&Exploring the Design Space of Optical See-through AR Head-Mounted Displays to Support First Responders in the Field&CHI \\
 &\citep{metaversepaper120}&EITPose: Wearable and Practical Electrical Impedance Tomography for Continuous Hand Pose Estimation&CHI \\
 &\citep{metaversepaper121}&Assessing User Apprehensions About Mixed Reality Artifacts and Applications: The Mixed Reality Concerns (MRC) Questionnaire&CHI \\
 &\citep{metaversepaper122}&What You Experience is What We Collect: User Experience Based Fine-Grained Permissions for Everyday Augmented Reality&CHI \\
 &\citep{metaversepaper123}&Kinetic Signatures: A Systematic Investigation of Movement-Based User Identification in Virtual Reality&CHI \\
 &\citep{metaversepaper124}&Privacy in Immersive Extended Reality: Exploring User Perceptions, Concerns, and Coping Strategies&CHI \\
 &\citep{metaversepaper125}&An Empirical Study on Oculus Virtual Reality Applications: Security and Privacy Perspectives&ICSE \\
 &\citep{metaversepaper129}&EgoTouch: On-Body Touch Input Using AR/VR Headset Cameras&UIST \\
 &\citep{metaversepaper130}&Motion Passwords&VR \\
 &\citep{metaversepaper133}&A Generative Framework for Low-Cost Result Validation of Machine Learning-as-a-Service Inference&AsiaCCS \\
 &\citep{metaversepaper135}&<italic>xr-droid</italic>: A Benchmark Dataset for AR/VR and Security Applications&TODSC \\
 &\citep{metaversepaper136}&Time to Think the Security of WiFi-Based Behavior Recognition Systems&TODSC \\
 &\citep{metaversepaper137}&Dangers Behind Charging VR Devices: Hidden Side Channel Attacks via Charging Cables&TOIFS \\
 &\citep{metaversepaper138}&An Anti-Disguise Authentication System Using the First Impression of Avatar in Metaverse&TOIFS \\
 &\citep{metaversepaper139}&Privacy-Preserving Gaze Data Streaming in Immersive Interactive Virtual Reality: Robustness and User Experience&TVCG \\
 &\citep{metaversepaper140}&Berkeley Open Extended Reality Recordings 2023 (BOXRR-23): 4.7 Million Motion Capture Recordings from 105,000 XR Users&TVCG \\
 &\citep{metaversepaper141}&Analysis and Design of Efficient Authentication Techniques for Password Entry with the Qwerty Keyboard for VR Environments&TVCG \\
 &\citep{metaversepaper142}&Usable Authentication in Virtual Reality: Exploring the Usability of PINs and Gestures&ACNS \\
 &\citep{metaversepaper143}&Leveraging Overshadowing for Time-Delay Attacks in 4G/5G Cellular Networks: An Empirical Assessment&ARES \\
 &\citep{metaversepaper145}&Securing Contrastive mmWave-based Human Activity Recognition against Adversarial Label Flipping&WiSec \\
 &\citep{metaversepaper146}&De-anonymizing VR Avatars using Non-VR Motion Side-channels&WiSec \\[4pt]
2023&\citep{metaversepaper019}&LocIn: Inferring Semantic Location from Spatial Maps in Mixed Reality&USS \\
 &\citep{metaversepaper021}&Erebus: Access Control for Augmented Reality Systems&USS \\
 &\citep{metaversepaper023}&Going through the motions: AR/VR keylogging from user head motions&USS \\
 &\citep{metaversepaper088}&Reframe: An Augmented Reality Storyboarding Tool for Character-Driven Analysis of Security \& Privacy Concerns&UIST \\
 &\citep{metaversepaper097}&Visual Indicators Representing Avatars' Authenticity in Social Virtual Reality and Their Impacts on Perceived Trustworthiness&TVCG \\
 &\citep{metaversepaper022}&Unique Identification of 50, 000+ Virtual Reality Users from Head \& Hand Motion Data&USS \\
 &\citep{metaversepaper024}&Is Your Wallet Snitching On You? An Analysis on the Privacy Implications of Web3&USS \\
 &\citep{metaversepaper025}&It's all in your head(set): Side-channel attacks on AR/VR systems&USS \\
 &\citep{metaversepaper060}&A Tagging Solution to Discover IoT Devices in Apartments&ACSAC \\
 &\citep{metaversepaper042}&Exploring the Privacy Risks of Adversarial VR Game Design&PETS \\
 &\citep{metaversepaper043}&Speculative Privacy Concerns About AR Glasses Data Collection&PETS \\
 &\citep{metaversepaper049}&Investigating Security Indicators for Hyperlinking Within the Metaverse&SOUPS \\
 &\citep{metaversepaper050}&An Investigation of Teenager Experiences in Social Virtual Reality from Teenagers', Parents', and Bystanders' Perspectives&SOUPS \\
 &\citep{metaversepaper056}&Virtual reality for improving cyber situational awareness in security operations centers&CS \\
 &\citep{metaversepaper098}&Understanding Context to Capture when Reconstructing Meaningful Spaces for Remote Instruction and Connecting in XR&CHI \\
 &\citep{metaversepaper018}&Exploring User Reactions and Mental Models Towards Perceptual Manipulation Attacks in Mixed Reality&USS \\
 &\citep{metaversepaper041}&Understanding Person Identification Through Gait&PETS \\
 &\citep{metaversepaper055}&Rise of the Metaverse?s Immersive Virtual Reality Malware and the Man-in-the-Room Attack \& Defenses&CS \\
 &\citep{metaversepaper062}&Eliciting Security \& Privacy-Informed Sharing Techniques for Multi-User Augmented Reality&CHI \\
 &\citep{metaversepaper044}&Securing Data Exchange in the Convergence of Metaverse and IoT Applications&ARES \\
 &\citep{metaversepaper007}&Privacy Leakage via Unrestricted Motion-Position Sensors in the Age of Virtual Reality: A Study of Snooping Typed Input on Virtual Keyboards&SP \\
 &\citep{metaversepaper008}&Low-effort VR Headset User Authentication Using Head-reverberated Sounds with Replay Resistance&SP \\
 &\citep{metaversepaper010}&SoundLock: A Novel User Authentication Scheme for VR Devices Using Auditory-Pupillary Response&NDSS \\
 &\citep{metaversepaper020}&Hidden Reality: Caution, Your Hand Gesture Inputs in the Immersive Virtual World are Visible to All!&USS \\
 &\citep{metaversepaper059}&FaceReader: Unobtrusively Mining Vital Signs and Vital Sign Embedded Sensitive Info via AR/VR Motion Sensors&CCS \\
 &\citep{metaversepaper086}&Going Incognito in the Metaverse: Achieving Theoretically Optimal Privacy-Usability Tradeoffs in VR&UIST \\
 &\citep{metaversepaper087}&SmartPoser: Arm Pose Estimation with a Smartphone and Smartwatch Using UWB and IMU Data&UIST \\
 &\citep{metaversepaper037}&A Secure Authentication Framework to Guarantee the Traceability of Avatars in Metaverse&TOIFS \\
 &\citep{metaversepaper096}&MagLoc-AR: Magnetic-Based Localization for Visual-Free Augmented Reality in Large-Scale Indoor Environments&TVCG \\
 &\citep{metaversepaper040}&SigA: RPPG-Based Authentication for Virtual Reality Head-Mounted Display&RAID \\[4pt]
2022&\citep{metaversepaper012}&OVRSEEN: Auditing Network Traffic and Privacy Policies in Oculus VR&USS \\
 &\citep{metaversepaper084}&NailRing: An Intelligent Ring for Recognizing Micro-gestures in Mixed Reality&ISMAR \\
 &\citep{metaversepaper039}&Hidden in Plain Sight: Exploring Privacy Risks of Mobile Augmented Reality Applications&TOPS \\
 &\citep{metaversepaper063}&Something Personal from the Metaverse: Goals, Topics, and Contextual Factors of Self-Disclosure in Commercial Social VR&CHI \\
 &\citep{metaversepaper099}&The Dark Side of Perceptual Manipulations in Virtual Reality&CHI \\
 &\citep{metaversepaper006}&SoK: Authentication in Augmented and Virtual Reality&SP \\
 &\citep{metaversepaper107}&Virtual Reality Observations: Using Virtual Reality to Augment Lab-Based Shoulder Surfing Research&VR \\
 &\citep{metaversepaper107}&Can I Borrow Your ATM? Using Virtual Reality for (Simulated) In Situ Authentication Research&VR \\
 &\citep{metaversepaper045}&SoK: A Systematic Literature Review of Knowledge-Based Authentication on Augmented Reality Head-Mounted Displays&ARES \\
 &\citep{metaversepaper085}&Enabling Customizable Workflows for Industrial AR Applications&ISMAR \\
 &\citep{metaversepaper070}&Combining Real-World Constraints on User Behavior with Deep Neural Networks for Virtual Reality (VR) Biometrics&VR \\
 &\citep{metaversepaper071}&HoloLogger: Keystroke Inference on Mixed Reality Head Mounted Displays&VR \\
 &\citep{metaversepaper072}&SPAA: Stealthy Projector-based Adversarial Attacks on Deep Image Classifiers&VR \\
 &\citep{metaversepaper073}&Temporal Effects in Motion Behavior for Virtual Reality (VR) Biometrics&VR \\
&\citep{metaversepaper074}&A Keylogging Inference Attack on Air-Tapping Keyboards in Virtual Environments&VR \\[4pt]
2021&\citep{metaversepaper067}&Covert Embodied Choice: Decision-Making and the Limits of Privacy Under Biometric Surveillance&CHI \\
 &\citep{metaversepaper068}&Crowdsourcing Design Guidance for Contextual Adaptation of Text Content in Augmented Reality&CHI \\
 &\citep{metaversepaper014}&AdCube: WebVR Ad Fraud and Practical Confinement of Third-Party Ads&USS \\
 &\citep{metaversepaper112}&ARENA: The Augmented Reality Edge Networking Architecture&ISMAR \\
 &\citep{metaversepaper065}&Understanding User Identification in Virtual Reality Through Behavioral Biometrics and the Effect of Body Normalization&CHI \\
 &\citep{metaversepaper066}&RepliCueAuth: Validating the Use of a Lab-Based Virtual Reality Setup for Evaluating Authentication Systems&CHI \\
 &\citep{metaversepaper077}&Influence of Interactivity and Social Environments on User Experience and Social Acceptability in Virtual Reality&VR \\
 &\citep{metaversepaper033}&Immersive Virtual Reality Attacks and the Human Joystick&TODSC \\
 &\citep{metaversepaper013}&Kal epsilon ido: Real-Time Privacy Control for Eye-Tracking Systems&USS \\
 &\citep{metaversepaper075}&Using Siamese Neural Networks to Perform Cross-System Behavioral Authentication in Virtual Reality&VR \\
 &\citep{metaversepaper076}&VR-Spy: A Side-Channel Attack on Virtual Key-Logging in VR Headsets&VR \\
 &\citep{metaversepaper093}&A privacy-preserving approach to streaming eye-tracking data&TVCG \\
 &\citep{metaversepaper051}&Spectrum-flexible secure broadcast ranging&WiSec \\
 &\citep{metaversepaper036}&Designing Leakage-Resilient Password Entry on Head-Mounted Smart Wearable Glass Devices&TOIFS \\
&\citep{metaversepaper028}&Global Feature Analysis and Comparative Evaluation of Freestyle In-Air-Handwriting Passcode for User Authentication&ACSAC \\
&\citep{metaversepaper048}&Evaluating anpd redefining smartphone permissions with contextualized justifications for mobile augmented reality apps&SOUPS \\[4pt]
2020&\citep{metaversepaper009}&OcuLock: Exploring Human Visual System for Authentication in Virtual Reality Head-mounted Display&NDSS \\
 &\citep{metaversepaper102}&The Security-Utility Trade-off for Iris Authentication and Eye Animation for Social Virtual Avatars&TVCG \\
 &\citep{metaversepaper038}&Mimicry Attacks on Smartphone Keystroke Authentication&TOPS \\[4pt]
2019&\citep{metaversepaper015}&Secure Multi-User Content Sharing for Augmented Reality Applications&USS \\
&\citep{metaversepaper078}&Investigating the Third Dimension for Authentication in Immersive Virtual Reality and in the Real World&VR \\
&\citep{metaversepaper079}&Person Independent, Privacy Preserving, and Real Time Assessment of Cognitive Load using Eye Tracking in a Virtual Reality Setup&VR \\
&\citep{metaversepaper089}&Designing AR Visualizations to Facilitate Stair Navigation for People with Low Vision&UIST \\
&\citep{metaversepaper095}&Collaborative Visual Analysis with Multi-level Information Sharing Using a Wall-Size Display and See-Through HMDs&PACIFICVIS \\
&\citep{metaversepaper104}&Behavioural Biometrics in VR: Identifying People from Body Motion and Relations in Virtual Reality&CHI \\
&\citep{metaversepaper029}&A First Look into Privacy Leakage in 3D Mixed Reality Data&ESORICS \\
&\citep{metaversepaper032}&GaitLock: Protect Virtual and Augmented Reality Headsets Using Gait&TODSC \\[4pt]
2018&\citep{metaversepaper005}&Towards Security and Privacy for Multi-user Augmented Reality: Foundations with End Users&SP \\
&\citep{metaversepaper081}&Human Identification Using Neural Network-Based Classification of Periodic Behaviors in Virtual Reality&VR \\
&\citep{metaversepaper047}&Ethics emerging: The story of privacy and security perceptions in virtual reality&SOUPS \\[4pt]
2017&\citep{metaversepaper004}&Securing Augmented Reality Output&SP \\
&\citep{metaversepaper027}&HoloPair: Securing Shared Augmented Reality Using Microsoft HoloLens&ACSAC \\
&\citep{metaversepaper052}&Visualizing the New Zealand Cyber Security Challenge for Attack Behaviors&TrustCom \\[4pt]
2016&\citep{metaversepaper003}&Prepose: Privacy, Security, and Reliability for Gesture-Based Programming&SP \\
&\citep{metaversepaper094}&Efficient verification of holograms using mobile augmented reality&TVCG \\
&\citep{metaversepaper016}&Virtual U: Defeating Face Liveness Detection by Building Virtual Models From Your Public Photos&USS \\[4pt]
2015&\citep{metaversepaper002}&SurroundWeb: Mitigating Privacy Concerns in a 3D Web Browser&SP \\
&\citep{metaversepaper090}&Candid Interaction: Revealing Hidden Mobile and Wearable Computing Activities&UIST \\
&\citep{metaversepaper031}&Design and Analysis of Shoulder Surfing Resistant PIN Based Authentication Mechanisms on Google Glass&FC \\
&\citep{metaversepaper030}&Visual Cryptography and Obfuscation: A Use-Case for Decrypting and Deobfuscating Information Using Augmented Reality&FC \\[4pt]
2014&\citep{metaversepaper069}&In situ with bystanders of augmented reality glasses: perspectives on recording and privacy-mediating technologies&CHI \\
&\citep{metaversepaper001}&World-Driven Access Control for Continuous Sensing&CCS \\
   \hline
\end{tabular}
\end{table*}

Table~\ref{tab:metaversepaper2} shows a summary of the 114 articles included in the study. 
We notice that most studies are spread across evaluation (40), technique (39) and application (26) categories, with only a handful focused on models (5) and systems (4).
It appears that security and privacy research in the metaverse is more practically oriented, focusing on impact rather than creating new conceptual frameworks, architectures, or theoretical advancements.
\ins{This practical orientation may signal a shift toward real-world applicability, suggesting a gradual maturation of metaverse security and privacy research.
The observed predominance of authentication-focused studies suggests that the metaverse research community is still addressing foundational challenges related to user verification and secure access in immersive environments. This emphasis reflects the early-stage nature of many XR platforms, where reliable identification mechanisms are a prerequisite for trust, personalization, and safety. In addition, the growing number of practical evaluations, especially user studies and benchmarking, indicates a shift from conceptual exploration to applied research that tests real-world usability, effectiveness, and security results. This trend toward empirical validation may signify the maturation of the field, as researchers increasingly prioritize deployable solutions over theoretical models. It also underscores the need for methodological rigor and cross-disciplinary collaboration to ensure that emerging systems are not only secure, but also usable, inclusive, and adaptable to diverse contexts of adoption.}

Our findings indicate that authentication (22) and unobservability (14) are the most frequently studied security and privacy properties, respectively.
They are frequently observed in multiple techniques (11 and 8) and evaluations (7 and 4), but are rarely seen in applications (2 and 1).
This suggests that research is currently focusing on broader aspects of authentication and unobservability rather than specific issues, perhaps because these specific issues have yet to be identified.
We observe that research efforts are evenly distributed across metaverse components such as virtual worlds (74), lifelogging (56), and augmented reality (55); this is consistent with virtual (66) and augmented (55) reality being the terms most frequently encountered in studies.
\ins{The relatively even distribution of research across metaverse components, such as virtual worlds, lifelogging, and augmented reality, suggests a broad and exploratory phase in the development of metaverse security and privacy research. Rather than coalescing around a dominant platform or technology, the field is actively investigating risks and design considerations across a variety of immersive paradigms. This diversity may reflect the fragmented nature of current metaverse technologies, where no single platform or interaction model has yet emerged as a clear standard. It also highlights the importance of tailoring security and privacy solutions to the specific advantages and vulnerabilities of each component, for example, continuous sensing in AR, persistent data collection in lifelogging, or identity persistence in virtual worlds. As the metaverse continues to evolve, this balanced research landscape provides a foundation for comparative studies and the eventual development of cross-platform security frameworks that can accommodate heterogeneous environments.}
All \chg{the}{technique} articles \del{on techniques} reported the use of benchmarking research strategies, and some (8) also describe methods that involve the evaluation of user performance and experience, which shows the need for comprehensive evaluations.
\ins{The exclusive reliance on benchmarking in technique-oriented papers, with only a minority incorporating user performance or experience evaluations, highlights a gap in how technical contributions are validated in metaverse security and privacy research. This suggests a need for more comprehensive evaluation strategies that not only measure technical efficiency but also consider usability, user perception, and real-world applicability, factors that are crucial in immersive and interactive systems.}
Both evaluation and application papers place a strong emphasis on users.
Evaluations integrate a combination of human experimentation (41), interviews (26), and questionnaires (24).
This aligns with the numerous applications and techniques focused on optimizing algorithm performance, as well as evaluations focused on user performance and experience.
We also identified several studies (50) that evaluated the environment and work practices, reflecting a pragmatic approach focused on grasping the specific requirements when integrating metaverse technologies to meet security and privacy demands.
\begin{table*}[ht]
\centering
\footnotesize

\renewcommand\arraystretch{0.1}
\caption{The number of papers by ranked venues.}

\label{tab:metaversepaper2}
\setlength\tabcolsep{-2.5pt}
\begin{tabular}{p{1.6cm}MMMMcMMMMcKKKKKKKcKKKKKKcLLLLLLLcNNNNcc}
    &
    
    \multicolumn{1}{c}{\parbox[t]{3mm}{\rotatebox[origin=l]{90}{Virtual World}}}& \multicolumn{1}{c}{\parbox[t]{3mm}{\rotatebox[origin=l]{90}{Lifelogging}}}& \multicolumn{1}{c}{\parbox[t]{3mm}{\rotatebox[origin=l]{90}{Augmented Reality}}}& \multicolumn{1}{c}{\parbox[t]{3mm}{\rotatebox[origin=l]{90}{Mirror World}}}&\multicolumn{1}{c}{\parbox[t]{3mm}{\rotatebox[origin=l]{90}{}}}& 

     \multicolumn{1}{c}{\parbox[t]{3mm}{\rotatebox[origin=l]{90}{Virtual Reality}}}& \multicolumn{1}{c}{\parbox[t]{3mm}{\rotatebox[origin=l]{90}{Augmented Reality}}}& \multicolumn{1}{c}{\parbox[t]{3mm}{\rotatebox[origin=l]{90}{Mixed Reality}}}& \multicolumn{1}{c}{\parbox[t]{3mm}{\rotatebox[origin=l]{90}{Extended Reality}}}&\multicolumn{1}{c}{\parbox[t]{3mm}{\rotatebox[origin=l]{90}{}}}& 
    
    \multicolumn{1}{c}{\parbox[t]{3mm}{\rotatebox[origin=l]{90}{Authentication}}} & \multicolumn{1}{c}{\parbox[t]{3mm}{\rotatebox[origin=l]{90}{Confidentiality}}} & \multicolumn{1}{c}{\parbox[t]{3mm}{\rotatebox[origin=l]{90}{Authorization}}} & \multicolumn{1}{c}{\parbox[t]{3mm}{\rotatebox[origin=l]{90}{Integrity}}}& \multicolumn{1}{c}{\parbox[t]{3mm}{\rotatebox[origin=l]{90}{Identification}}} & \multicolumn{1}{c}{\parbox[t]{3mm}{\rotatebox[origin=l]{90}{Availability}}} & \multicolumn{1}{c}{\parbox[t]{3mm}{\rotatebox[origin=l]{90}{Non-repudiation}}} & \multicolumn{1}{c}{\parbox[t]{3mm}{\rotatebox[origin=l]{90}{}}}&
 
    \multicolumn{1}{c}{\parbox[t]{3mm}{\rotatebox[origin=l]{90}{Unobservability}}} & \multicolumn{1}{c}{\parbox[t]{3mm}{\rotatebox[origin=l]{90}{Content Awareness}}} & \multicolumn{1}{c}{\parbox[t]{3mm}{\rotatebox[origin=l]{90}{Anonymity}}} & \multicolumn{1}{c}{\parbox[t]{3mm}{\rotatebox[origin=l]{90}{Policy}}} & \multicolumn{1}{c}{\parbox[t]{3mm}{\rotatebox[origin=l]{90}{Unlinkability}}} & \multicolumn{1}{c}{\parbox[t]{3mm}{\rotatebox[origin=l]{90}{Deniability}}} & \multicolumn{1}{c}{\parbox[t]{3mm}{\rotatebox[origin=l]{90}{}}}&
  
    \multicolumn{1}{c}{\parbox[t]{3mm}{\rotatebox[origin=l]{90}{Benchmarking}}} & \multicolumn{1}{c}{\parbox[t]{3mm}{\rotatebox[origin=l]{90}{Human Experimentation}}} & \multicolumn{1}{c}{\parbox[t]{3mm}{\rotatebox[origin=l]{90}{Interview}}}& \multicolumn{1}{c}{\parbox[t]{3mm}{\rotatebox[origin=l]{90}{Questionnaire}}}& \multicolumn{1}{c}{\parbox[t]{3mm}{\rotatebox[origin=l]{90}{Data Science}}}& \multicolumn{1}{c}{\parbox[t]{3mm}{\rotatebox[origin=l]{90}{Engineering Research}}}& \multicolumn{1}{c}{\parbox[t]{3mm}{\rotatebox[origin=l]{90}{Meta Science}}}& \multicolumn{1}{c}{\parbox[t]{3mm}{\rotatebox[origin=l]{90}{}}}& 

  \multicolumn{1}{c}{\parbox[t]{3mm}{\rotatebox[origin=l]{90}{Algorithm Performance}}}& \multicolumn{1}{c}{\parbox[t]{3mm}{\rotatebox[origin=l]{90}{Environment and Practices}}}& \multicolumn{1}{c}{\parbox[t]{3mm}{\rotatebox[origin=l]{90}{User Experience}}}& \multicolumn{1}{c}{\parbox[t]{3mm}{\rotatebox[origin=l]{90}{User Performance}}}&\multicolumn{1}{c}{\parbox[t]{3mm}{\rotatebox[origin=l]{90}{}}}&\\
  
\textbf{Type} & \multicolumn{5}{c}{\textbf{Metaverse}} & \multicolumn{5}{c}{\textbf{Reality}} & \multicolumn{8}{c}{\textbf{Security}} & \multicolumn{7}{c}{\textbf{Privacy}} & \multicolumn{8}{c}{\textbf{Strategy}} & \multicolumn{5}{c}{\textbf{Evaluation}}& \textbf{Total} \\
 
 \hline
    \textbf{Application}&15&11&15&6&&13&15&4&3&&2&3&4&3&2&1&2&&1&2&1&3&1&1&&14&14&2&9&4&4&0&&16&11&7&7&&\textbf{26}\\[3pt]
    \textbf{Evaluation}&27&13&16&7&&24&16&3&3&&7&5&3&1&4&0&0&&4&10&2&3&1&0&&5&27&24&15&4&2&1&&6&31&15&13&&\textbf{40}\\[3pt]
    \textbf{Model}&3&2&2&3&&2&2&1&1&&2&0&0&1&1&0&0&&1&0&0&0&0&0&&4&0&0&1&2&1&2&&4&1&0&0&&\textbf{5}\\[3pt]
    \textbf{System}&1&2&3&2&&0&3&1&1&&0&1&2&0&0&0&0&&0&0&0&0&1&0&&3&0&1&0&0&1&0&&2&2&0&0&&\textbf{4}\\[3pt]
    \textbf{Technique}&28&28&19&12&&27&19&7&2&&11&3&1&5&2&2&0&&8&0&5&0&1&1&&39&16&3&4&16&1&0&&39&5&6&8&&\textbf{39}\\[3pt]

    \hline
    
    \textbf{Total}&\multicolumn{1}{T}{74}&\multicolumn{1}{T}{56}&\multicolumn{1}{T}{55}&\multicolumn{1}{T}{30}&&\multicolumn{1}{T}{66}&\multicolumn{1}{T}{55}&\multicolumn{1}{T}{16}&\multicolumn{1}{T}{10}&&\multicolumn{1}{T}{22}&\multicolumn{1}{T}{12}&\multicolumn{1}{T}{10}&\multicolumn{1}{T}{10}&\multicolumn{1}{T}{9}&\multicolumn{1}{T}{3}&\multicolumn{1}{T}{2}&&\multicolumn{1}{T}{14}&\multicolumn{1}{T}{12}&\multicolumn{1}{T}{8}&\multicolumn{1}{T}{6}&\multicolumn{1}{T}{4}&\multicolumn{1}{T}{2}&&\multicolumn{1}{T}{65}&\multicolumn{1}{T}{57}&\multicolumn{1}{T}{30}&\multicolumn{1}{T}{29}&\multicolumn{1}{T}{26}&\multicolumn{1}{T}{9}&\multicolumn{1}{T}{3}&&\multicolumn{1}{T}{67}&\multicolumn{1}{T}{50}&\multicolumn{1}{T}{28}&\multicolumn{1}{T}{28}&&\textbf{114}\\[3pt]
    
    \\\\
    \hline
\end{tabular}
\end{table*}

Table~\ref{tab:metaversetable2} presents a more detailed list of the 26 venues in which included studies are published.\footnote{We investigated a total of 68 conferences and journals each year, from 2013 to 2024.} 
We notice that SE publications are only in one venue and HCI publications are largely confined to only six venues, in contrast to SP papers, which are spread across 19 venues.
Possibly due to the rapid progression of the metaverse, we observe that most studies (\ie 82\%) are published at conferences (which offer expedited review processes) rather than through journals to facilitate timely dissemination of findings.
\begin{table}[h!]
\centering
\tiny
\caption{Selected Publication Venues.}
\label{tab:metaversetable2}
\setlength\tabcolsep{1pt}
\begin{tabular}{llr}
    \hline
    \textbf{No} & \textbf{Name} & \textbf{Included} \\
    \hline
    1&IEEE Transactions on Visualization and Computer Graphics&8\\
    2&IEEE Transactions on Dependable and Secure Computing&4\\
    3&IEEE Transactions on Information Forensics and Security&4\\
    4&ACM Transactions on Privacy and Security&2\\
    5&Computers and Security&2\\[3pt]
    
    6&Usenix Security Symposium&17\\
    7&International Conference on Human Factors in Computing Systems&16\\
    8&IEEE Conference on Virtual Reality and 3D User Interfaces&14\\
    9&IEEE Symposium on Security and Privacy&8\\
    10&ACM Symposium on User Interface Software and Technology&6\\
    11&Symposium On Usable Privacy and Security&4\\
    12&IEEE/ACM International Symposium on Mixed and Augmented Reality&3\\
    13&Usenix Network and Distributed System Security Symposium&3\\
    14&Privacy Enhancing Technologies Symposium&3\\
    15&Annual Computer Security Applications Conference&3\\
    16&International Conference on Availability, Reliability and Security&3\\
    17&ACM Conference on Security and Privacy in Wireless and Mobile Networks&3\\
    18&ACM Conference on Computer and Communications Seurity&2\\
    19&Financial Cryptography and Data Security Conference&2\\
    20&IEEE Pacific Visualization Symposium&1\\
    21&European Symposium on Research in Computer Security&1\\
    22&Asia Conference on Information, Computer and Communications Security&1\\
    23&The International Symposium on Research in Attacks, Intrusions and Defenses&1\\
    24&International Conference on Trust, Security and Privacy in Computing and Communications&1\\
    25&International Conference on Applied Cryptography and Network Security&1\\
    26&International Conference on Software Engineering&1\\
    
    \hline
    \multicolumn{2}{r}{\textbf{Total}}&114\\
   \hline
\end{tabular}
\end{table}
\begin{rmd}
\begin{itemize}[leftmargin=1.5em, labelsep=0.5em, itemsep=0.5em]
    \item In the last decade, security and privacy risks in metaverse publications increased from 2 to 27 \ins{yearly}, evidencing greater attention to community research.
    \item Conference publications are almost five times more than journal publications (\ie 20 and 94, respectively), which could be attributed to a quick evaluation process in conferences aligning with the rapid advancement of the metaverse.
    \item The 44\% of the studies evaluated the environment and work practices, suggesting a practical approach to address security and privacy issues in the metaverse.
\end{itemize}
\end{rmd}
\subsection{Metaverse Focus} 
\begin{figure}[h!]
   \centering
   \includegraphics[width=0.48\textwidth]{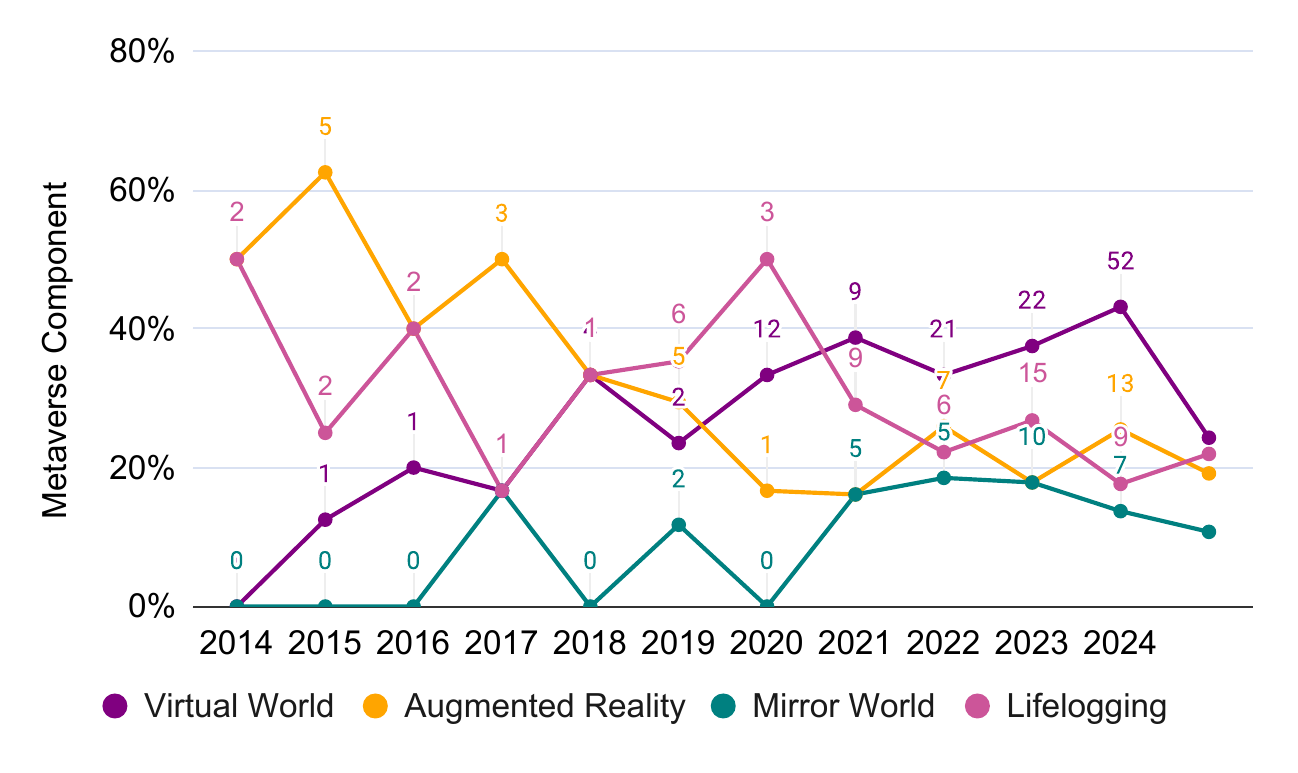}
   \caption{The evolution of metaverse components.}
   \label{fig:MetaverComponentEvolution}
\end{figure}

\begin{figure}[h]
   \centering
   \includegraphics[width=0.48\textwidth]{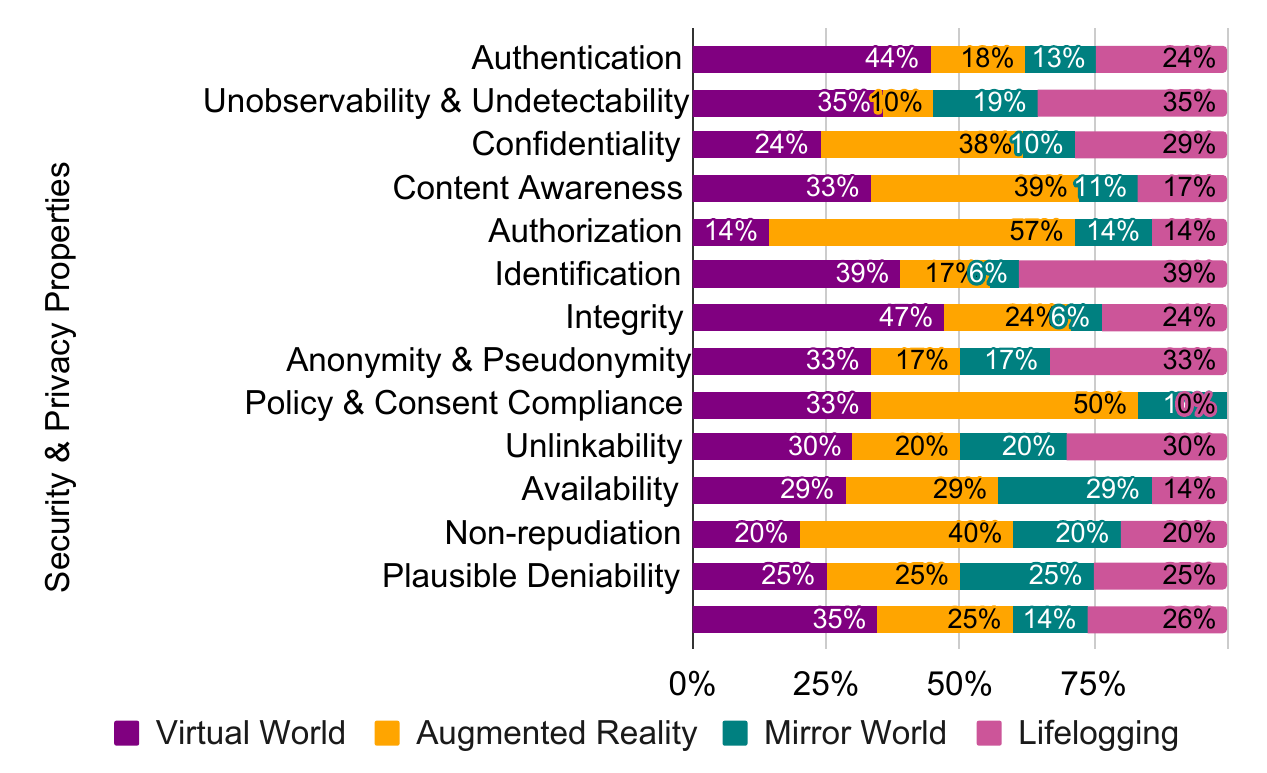}
   \caption{The distribution of metaverse components based on security and privacy properties.}
\label{fig:SPPropertiesByMetaverseComponents}
\end{figure}

Figure~\ref{fig:MetaverComponentEvolution} displays the results related to the components of the metaverse.
Each colored line corresponds to a specific metaverse component.
The $Y$ axis shows the percentage associated with each component, and the circles are marked with the count of studies involving those components.
In particular, there is a consistent decline in augmented reality, while virtual worlds show a steady rise.
The total number of studies has increased significantly in recent years, and in the last two years, the percentages of different components have converged.
In the past, the total number of studies was relatively small, with a higher percentage using AR probably due to smartphones being more accessible than VR headsets. Today, VR headsets have gained popularity and immersive AR headsets are available, although they are quite expensive. We anticipate that future research will focus on integrating various elements, as devices such as the Meta Quest 3 have made AR and VR more economically accessible.
Figure~\ref{fig:SPPropertiesByMetaverseComponents} illustrates a stacked bar chart that shows the distribution of metaverse components engaged in evaluating various security and privacy properties.
Most of these evaluations (\ie 42\%) focus on a single metaverse component.
However, some (\ie 37\%) involve two, and a small fraction (\ie 12\%) incorporate three components.
Properties like integrity, authentication, and unobservability often engage virtual worlds, whereas authentication, authorization, confidentiality, content awareness are associated with augmented reality.

Mirror worlds appear infrequently in evaluations, whereas lifelogging is more prominent in properties related to unobservability \& undetectability, and identification.
Possibly, complex social and economic interactions through avatars~\citep{metaversepaper007,metaversepaper086} in virtual worlds promote the investigation of security properties such as authentication and data integrity. However, the need for advanced mapping technologies and geospatial data~\citep{metaversepaper019} to develop mirror worlds makes it more challenging to incorporate them into security and privacy experiments. These factors could contribute to the fact that mirror worlds are less frequently included in evaluations of security and privacy properties compared to other metaverse components.

\begin{rmd}
\begin{itemize}[leftmargin=1.5em, labelsep=0.5em, itemsep=0.5em]
    \item In the last two years, the number of publications that involve virtual worlds (\ie 24\%), lifelogging (\ie 22\%), augmented reality (\ie 19\%), and mirror worlds (\ie 11\%), has almost converged.
    \item In 42\% of the cases, the evaluations focus on one component of the metaverse. 
    \item Typically, virtual worlds are evaluated for properties such as authentication (\ie 27\%), unobservability (\ie 15\%) and integrity (\ie 11\%). Lifelogging is often evaluated based on authentication (\ie 20\%), unobservability (\ie 20\%) and identification (\ie 13\%). Augmented reality tends to be analyzed in terms of authentication (\ie 15\%), confidentiality (\ie 15\%), and authorization (\ie 15\%). Mirror worlds evaluations usually involve authentication (\ie 20\%), unobservability (\ie 20\%), and anonymity (\ie 10\%).
\end{itemize}
\end{rmd}

There is an interesting observation in \del{the }authorization. We can see that 57\% of the studies focused on authorization involve AR, but only 14\% involve virtual worlds. This is likely due to differences in the interaction and environmental contexts. In AR, devices and software have a feature known as perceptual sensing. This refers to the ability of hardware or software to continuously monitor the physical environment using cameras and other sensors~\citep{metaversepaper001}.
Due to this capability, AR devices can unintentionally capture sensitive information, such as credit card numbers or the contents of computer screens. In addition, the current design of permission models in the underlying operating systems exacerbates the issue of overprivileged access~\citep{metaversepaper021}.

In the context of lifelogging, studies emphasize properties such as unobservability \& undetectability, as well as anonymity \& pseudonymity, with approximately 35\% and 33\% of articles addressing these properties, respectively.
This focus is understandable given the frequent use of pseudo-identities linked to user profiles for personalized services. However, properties such as policy compliance and consent management remain underexplored and pose significant challenges.
This difficulty arises from two main factors: regulatory and technical perspectives~\citep{wilkowska2023}. From a regulatory point of view, regulations such as the GDPR make it difficult to obtain explicit and unambiguous consent from data subjects before processing their data. From a technical perspective, obtaining fully informed consent is challenging because many devices lack screens, making it difficult to display privacy policies.

\subsection{Study Type}

\begin{figure}[h!]
   \centering
   \includegraphics[width=0.48\textwidth]{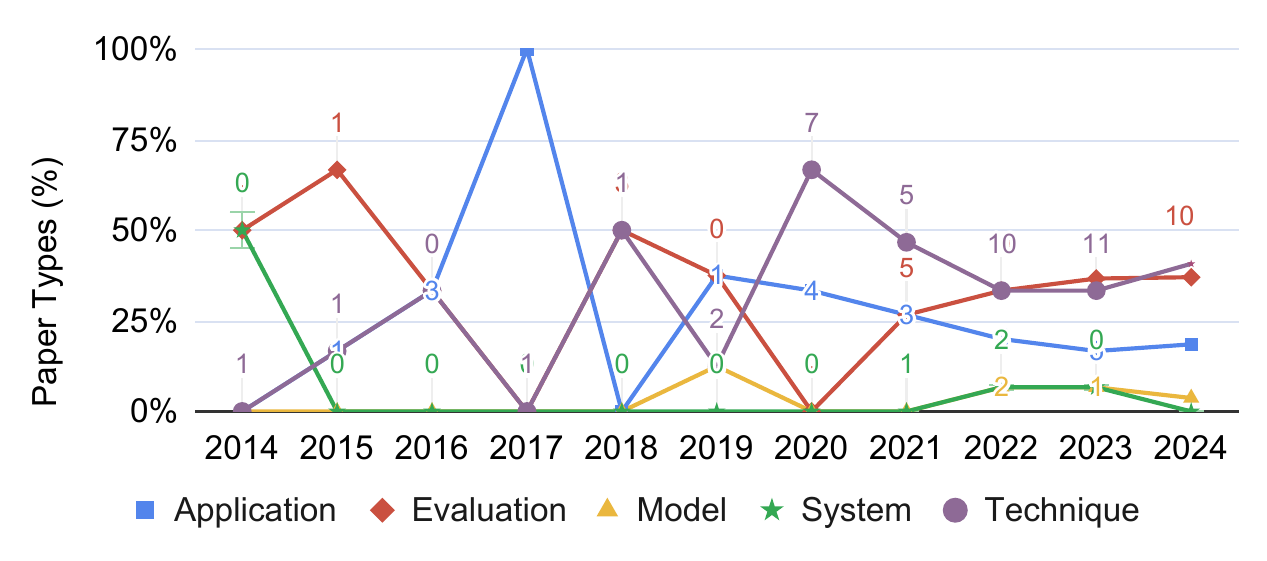}
   \caption{Trends of the types of papers per year.}
   \label{fig:PercetageOfPaperTypesByYear}
\end{figure}

Table~\ref{tab:metaversepaper2} presents a summary of the classification of papers by type. 
The colors highlight the dimensions being analyzed. 
A stronger color intensity indicates a larger number of papers discovered.
Evaluation and technique papers are the most common types. 
These types of paper often employ human experimentation (27) and benchmarking (39) methods.
It is important to note that human experimentation does not feature at all in model and system papers, highlighting the difficulties of engaging human participants in testing theories and developing frameworks.
Figure~\ref{fig:PercetageOfPaperTypesByYear} presents a line \chg{graph}{chart} with the trends in the percentage of paper types over time. 
The labels next to the marks indicate the absolute number of papers.
The patterns slightly mirror a funnel shape. Specifically, in the early years examined, there were considerable variations in all categories. 
However, in recent times, there has been a marked decrease in the percentage of applications and techniques, while evaluations have risen moderately and systems have experienced modest growth. 
In contrast, the absolute number of all types is increasing.
Papers on models are almost absent.
The emphasis may have shifted due to the urgent need for rapid advances in the metaverse and its inherent complexity~\citep{yunetwork2023}. 
This situation poses a challenge in creating models that remain pertinent and precise enough for comprehensive testing and validation, as foundational technology evolves quickly. 
In addition, the presence of software and immersive hardware allows for user assessments and the validation of existing methods' effectiveness.
\begin{rmd}
\begin{itemize}[leftmargin=1.5em, labelsep=0.5em, itemsep=0.5em]
    \item Although the total quantity in all types of paper continues to increase, indicating a vibrant landscape, the technique and evaluation paper have received more attention in recent years.
    \item The nearly lack of models found (\ie 4\%) emphasize the urgent need to develop a common understanding of the foundations to approach security and privacy concerns in the metaverse.
\end{itemize}
\end{rmd}

\subsection{Security \& Privacy Property}
\begin{figure}[!h]
   \centering
   \includegraphics[width=0.49\textwidth]{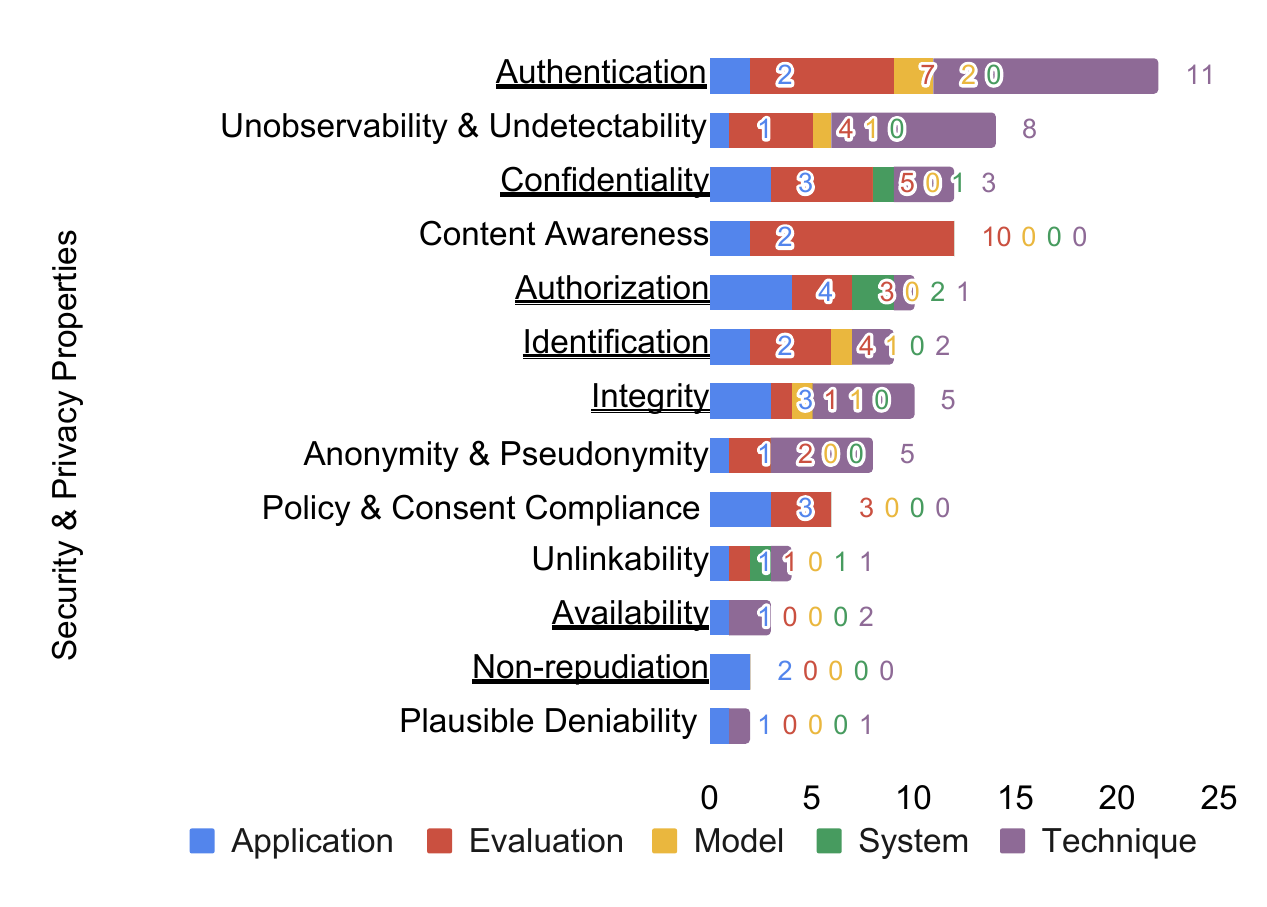}
   \caption{Security \& privacy properties involved in metaverse evaluations by paper types. The labels on security properties are underlined.}
   \label{fig:SPPropertiesByPaperType}
\end{figure}

Figure~\ref{fig:SPPropertiesByPaperType} presents a stacked bar chart with the total number of paper types by security and privacy properties. 
We found a small but consistent number of application papers for all security \& privacy properties with the exception of unobservability \& undetectability and identification.
Real-world usage contexts and direct interactions with users may influence properties like unobservability \& undetectability, and identification. Evaluating these properties requires empirical data, which can only be obtained through practical scenarios such as user studies, surveys, or case studies. 
These strategies may provide insight into how these properties function under actual conditions. 
Therefore, it is not surprising that evaluation papers tend to explore these properties in greater depth. 
In contrast, application papers often lack the scope to conduct empirical studies or extensive surveys.
We observe that multiple techniques address authentication (11) and unobservability \& undetectability (8), integrity (5), and anonymity \& pseudonymity (5).
\ins{The prominence of authentication may suggest that security and privacy concerns persist in metaverse development.
In effect, the strong focus on authentication across the reviewed studies reflects the nascent and infrastructure-building stage of metaverse technology adoption. As platforms strive to attract broader user bases and support increasingly complex interactions, the ability to reliably identify and verify users becomes fundamental. Unlike traditional web or mobile ecosystems, metaverse environments require continuous and often multimodal authentication through biometrics, behavioral signals, or embodied interactions, which introduces both technical challenges and privacy risks. The emphasis of the research community on authentication likely mirrors industry priorities, where trust, security, and prevention of impersonation are immediate concerns in enabling social, commercial, and enterprise use cases. This also suggests that before metaverse technologies can evolve towards richer, large-scale, and interoperable systems, the field must first establish robust identity frameworks that account for the unique characteristics of immersive, embodied interaction.}
Authentication techniques are essential to ensure that system or application access is restricted to verified users, thus preventing unauthorized entry. 
Thus, authentication is a vital attribute in the research on techniques that protect data and transactions in metaverse applications. 
For example, \emph{SigA}~\citep{metaversepaper040} is an innovative technique that uses a physiological signal that is invisible to the naked eye (photoplethysmogram) rather than the more commonly used electro-oculogram and electrical muscle stimulation methods for authentication. 
The technique improves security by reducing the risk of shoulder surfing attacks and strengthens user privacy by mitigating the threats posed by side-channel attacks.
\ins{In addition to security and privacy, it is also worth to consider authentication from other perspectives, including deployability, usability, and accessibility.}

We identified several papers (12) that address content awareness; however, none are classified as technique-type papers.
Content awareness is often discussed in studies that emphasize user interaction, such as evaluation or application.
In contrast, technique papers generally focus on solutions to specific issues, such as security or privacy-enhancing algorithms, without paying much attention to user experience. 
Consequently, aspects such as transparency and user awareness regarding the data they share are frequently neglected. 
\ins{The absence of content awareness in technique-type papers indicates a disconnect between technical solutions and user-facing concerns such as transparency and informed consent. This suggests that technical research on metaverse security may overlook critical aspects of user experience, potentially limiting the trustworthiness and adoption of proposed solutions. Bridging this gap requires integrating user-centered principles into the design and evaluation of security and privacy-enhancing technologies.}

We found that most evaluations focus on content awareness (10), authentication (7), confidentiality (5), unobservability \& undetectability (4), and identification (4).
We did not find evaluations that focus on properties such as availability, non-repudiation, and plausible deniability.
Limited research on, for example, anonymity and pseudonymity can be attributed to the complexity of privacy issues~\citep{liu2023}. 
One contributing factor to this complexity is the low level of user awareness~\citep{metaversepaper099}. 
Awareness typically develops over time and is influenced by users' experiences with the technology itself.
Without a proper understanding of privacy risks, users often underestimate the importance of privacy protection features in the metaverse. 
Therefore, two important aspects must be addressed: the development of technology that leverages user privacy and the need to raise user awareness about the risks present in the metaverse.
\begin{figure}[h!]
   \centering
   \includegraphics[width=0.49\textwidth]{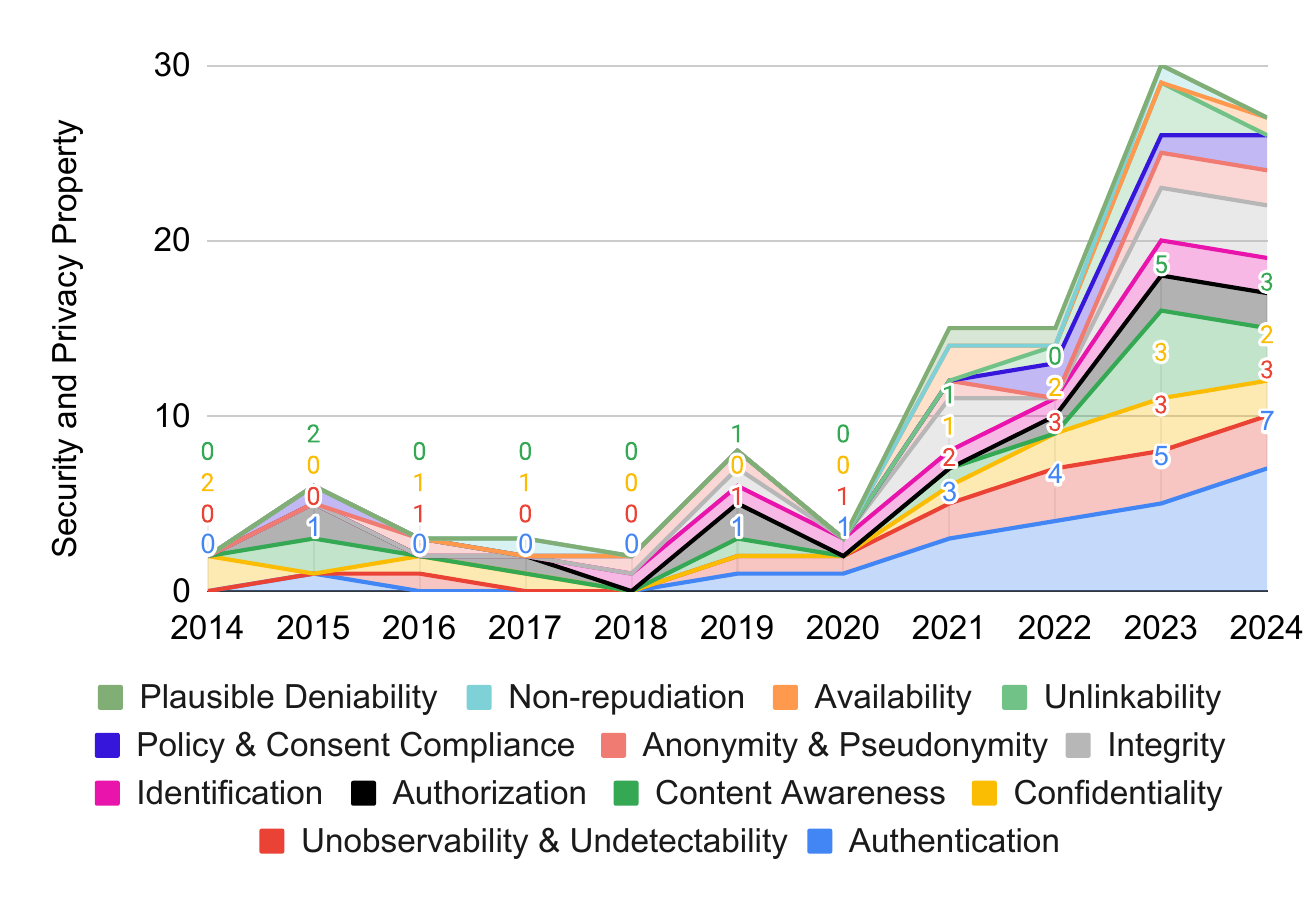}
   \caption{The evolution of security \& privacy properties between 2014 and 2024.}
   \label{fig:SPPropertiesEvolution}
\end{figure}

Figure~\ref{fig:SPPropertiesEvolution} presents a stacked area chart with the evolution of the number of articles by security and privacy properties.
Regardless of specific properties, there has been a steady increase in the number of research papers on metaverse security and privacy.
In the past three years, there has been a notable increase in articles dealing with authentication, unobservability and undetectability, confidentiality, and content awareness. 
Similarly, this field is expanding as the number of relevant topics grows, highlighting a broad focus on various security and privacy properties.   
In addition, there has been a notable increase in privacy-focused articles in the last three years.
\begin{rmd}
\begin{itemize}[leftmargin=1.5em, labelsep=0.5em, itemsep=0.5em]
    \item Authentication and confidentiality are the most common security properties (\ie 22 and 12 of 68 articles), while unobservability and content awareness are the most common privacy properties (\ie 14 and 12 of 46 articles).
    \item The field of security and privacy in the metaverse is growing, with research increasingly diversifying into a larger number of security and privacy properties over time.
\end{itemize}
\end{rmd}

\subsection{Research Strategy}
\begin{figure}[t]
   \centering
   \includegraphics[width=0.48\textwidth]{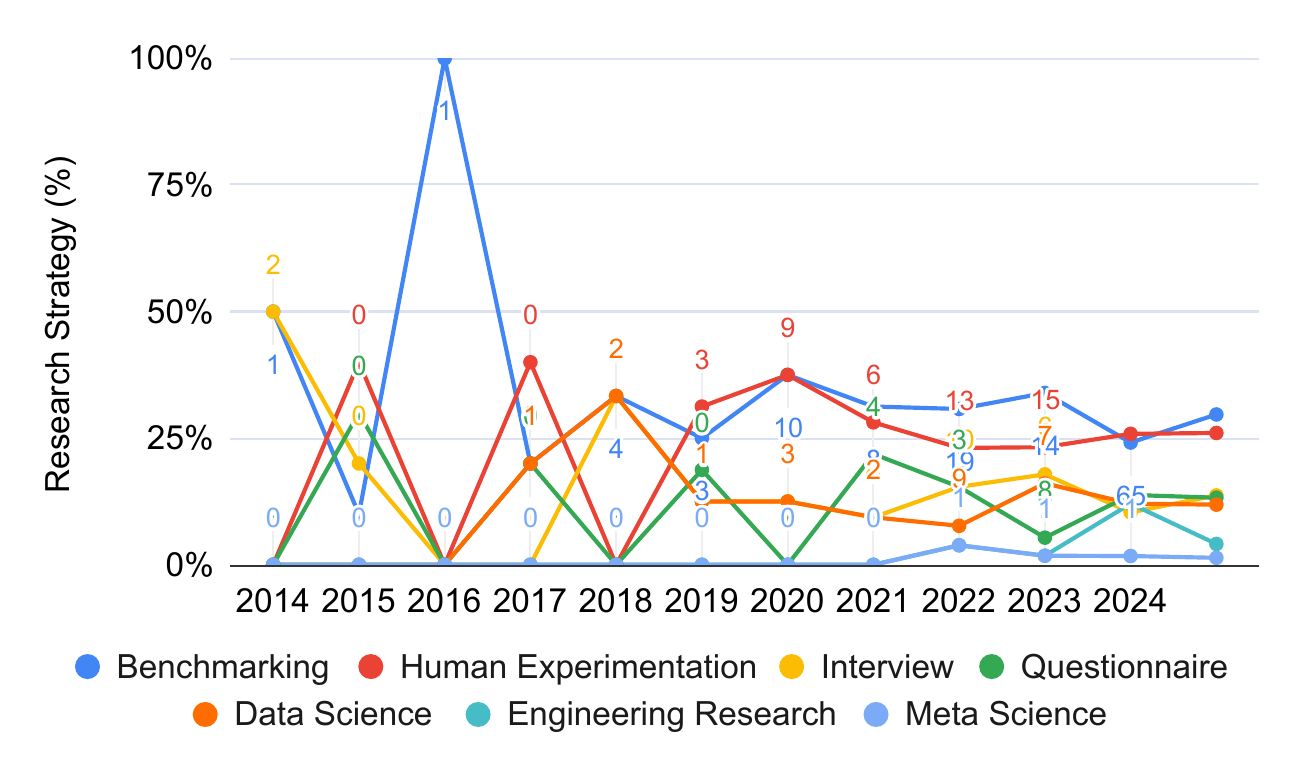}
   \caption{The evolution of research strategies used in the evaluation of security and privacy within the metaverse.}
   \label{fig:ResearchStrategiesEvaluation}
\end{figure}

Figure~\ref{fig:ResearchStrategiesEvaluation} illustrates the evolution of research strategies used to assess aspects of security and privacy within the metaverse.
Each colored line presents the evolution of the percentage of studies related to a specific research strategy over a year. 
The labels on the marks display the total number of studies that employ the research strategy. 
Notice that we collected all research strategies described in the articles and often found investigations that involve more than one strategy.  
Approximately 29\% employ a single strategy, around 50\% incorporate two, and the remaining 21\% use three strategies.
These findings could correspond to the complexity of immersive environments, where numerous variables can influence user behavior, highlighting the need for various methods in evaluations.
There is a significant increase in the number of strategies included in the investigation (in line with the increasing number of studies), with strategies becoming more evenly distributed over the past two years.
\begin{rmd}
\begin{itemize}[leftmargin=1.5em, labelsep=0.5em, itemsep=0.5em]
    \item The top 3 research strategies used to study security and privacy in the metaverse include benchmarking, human experiments, and interviews, accounting for 30\%, 26\%, and 14\% of usage, respectively.
    \item 71\% percent of the studies involve more than one research strategy, highlighting the complexity of immersive environments where numerous variables can influence user behavior, underscoring the need for various evaluation methods.
\end{itemize}
\end{rmd}
\subsection{Evaluation Scope}
Table~\ref{tab:metaversepaper3} presents an exhaustive list of the scope of the evaluations in the 114 articles analyzed.
We collect various details of the evaluations, such as the scenarios in which they occur (\eg algorithm performance, environment and work practices, user experience and performance), the scope of the studies and information about available artifacts. 
We observe that most evaluations (\ie 67\%) include human subjects. 
The widespread integration of real-world datasets along with accessible application artifacts highlights the progress towards practical reliability. 
In general, these results suggest that the focus on user-centered methods guarantees the reliability and relevance of these technologies in actual settings.
\begin{table}[t]
\centering
\scriptsize

\renewcommand\arraystretch{0.1}
\caption{Evaluation and Quality Metrics.}

\label{tab:metaversepaper3} 
\setlength\tabcolsep{-1.5pt}
\begin{tabular}{p{3.5cm}XXXXcYYYYcZZ}
    &
    
    \multicolumn{1}{c}{\parbox[t]{3mm}{\rotatebox[origin=l]{90}{Algorithm Performance}}}& \multicolumn{1}{c}{\parbox[t]{3mm}{\rotatebox[origin=l]{90}{User Performance}}}& \multicolumn{1}{c}{\parbox[t]{3mm}{\rotatebox[origin=l]{90}{User Experience}}}& \multicolumn{1}{c}{\parbox[t]{3mm}{\rotatebox[origin=l]{90}{Environment and Practices}}}&\multicolumn{1}{c}{\parbox[t]{3mm}{\rotatebox[origin=l]{90}{}}}& 

     \multicolumn{1}{c}{\parbox[t]{3mm}{\rotatebox[origin=l]{90}{Correctness}}}& \multicolumn{1}{c}{\parbox[t]{3mm}{\rotatebox[origin=l]{90}{Usability}}}& \multicolumn{1}{c}{\parbox[t]{3mm}{\rotatebox[origin=l]{90}{Robustness}}}& \multicolumn{1}{c}{\parbox[t]{3mm}{\rotatebox[origin=l]{90}{Time}}}&\multicolumn{1}{c}{\parbox[t]{3mm}{\rotatebox[origin=l]{90}{}}}& 
    
    \multicolumn{1}{c}{\parbox[t]{3mm}{\rotatebox[origin=l]{90}{Average}}} & \multicolumn{1}{c}{\parbox[t]{3mm}{\rotatebox[origin=l]{90}{Std. Dev}}} \\
  
\textbf{Property} & \multicolumn{5}{c}{\textbf{Evaluation}} & \multicolumn{5}{c}{\textbf{Quality}} & \multicolumn{2}{c}{\textbf{Participants}}\\
 
 \hline

\textbf{Authentication}&13&9&8&8&&13&11&8&6&&37.0&36.0\\[3pt]
\textbf{Unobservability \& Undetect.}&12&1&4&3&&11&4&9&0&&15.1&14.3\\[3pt]
\textbf{Confidentiality}&7&3&3&6&&5&7&3&4&&26.5&33.8\\[3pt]
\textbf{Content Awareness}&0&3&5&12&&0&12&0&0&&59.0&70.1\\[3pt]
\textbf{Authorization}&5&1&1&5&&2&5&5&0&&4.8&8.0\\[3pt]
\textbf{Identification}&6&3&2&5&&5&3&5&1&&112.7&269.4\\[3pt]
\textbf{Integrity}&8&3&2&4&&7&2&4&5&&22.9&27.9\\[3pt]
\textbf{Anonymity \& Pseudonymity}&6&2&0&1&&5&1&3&0&&68.1&160.4\\[3pt]
\textbf{Policy \& Consent Compliance}&2&2&2&4&&1&3&2&0&&55.0&115.1\\[3pt]
\textbf{Unlinkability}&3&0&0&1&&2&2&1&2&&6.8&5.4\\[3pt]
\textbf{Availability}&3&0&0&0&&1&0&2&1&&266.7&461.9\\[3pt]
\textbf{Non-repudiation}&1&0&1&0&&1&1&0&0&&10.5&14.8\\[3pt]
\textbf{Plausible Deniability}&1&1&0&1&&1&0&1&1&&27.5&38.9\\[3pt]
    \hline
\textbf{Total}&\multicolumn{1}{c}{67}&\multicolumn{1}{c}{28}&\multicolumn{1}{c}{28}&\multicolumn{1}{c}{50}&&\multicolumn{1}{c}{55}&\multicolumn{1}{c}{51}&\multicolumn{1}{c}{44}&\multicolumn{1}{c}{20}&&\multicolumn{1}{c}{54.8}&\multicolumn{1}{c}{133.4}\\[3pt] 
    \\\\
    \hline
\end{tabular}
\end{table}

\begin{figure*}[ht!]
    \centering
    \includegraphics[width=\textwidth]
    {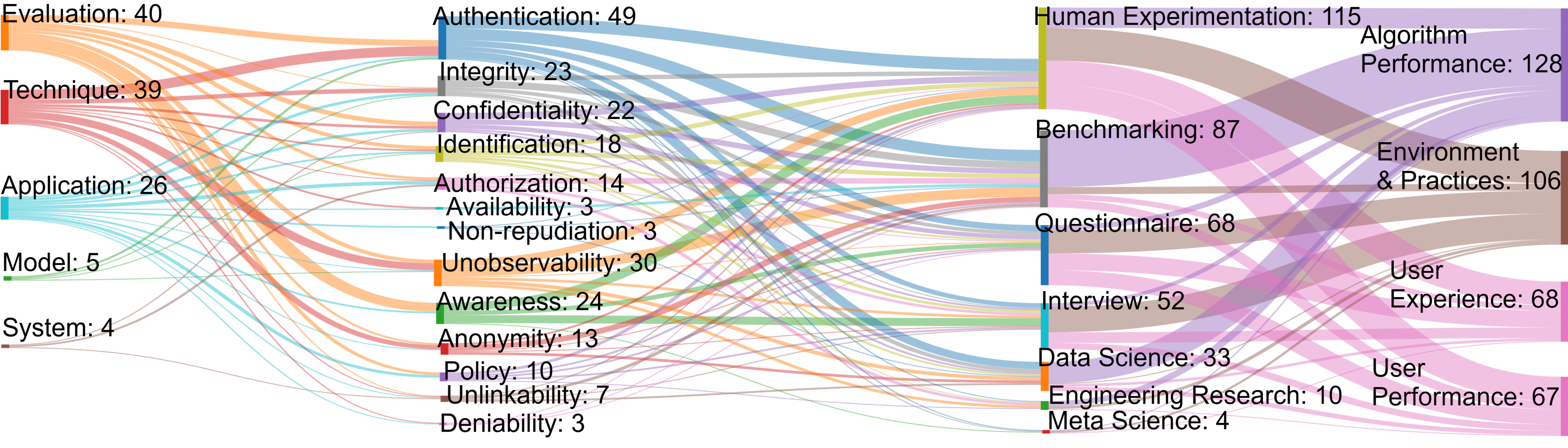}
    \caption{Relationship between evaluation scenarios, study types, venues, S\&P properties, type of reality.
    }
    \label{fig:sankeypapertypeandresearchtopic1a}
\end{figure*}

\begin{figure}[h!]
  \centering
   \includegraphics[width=0.48\textwidth]{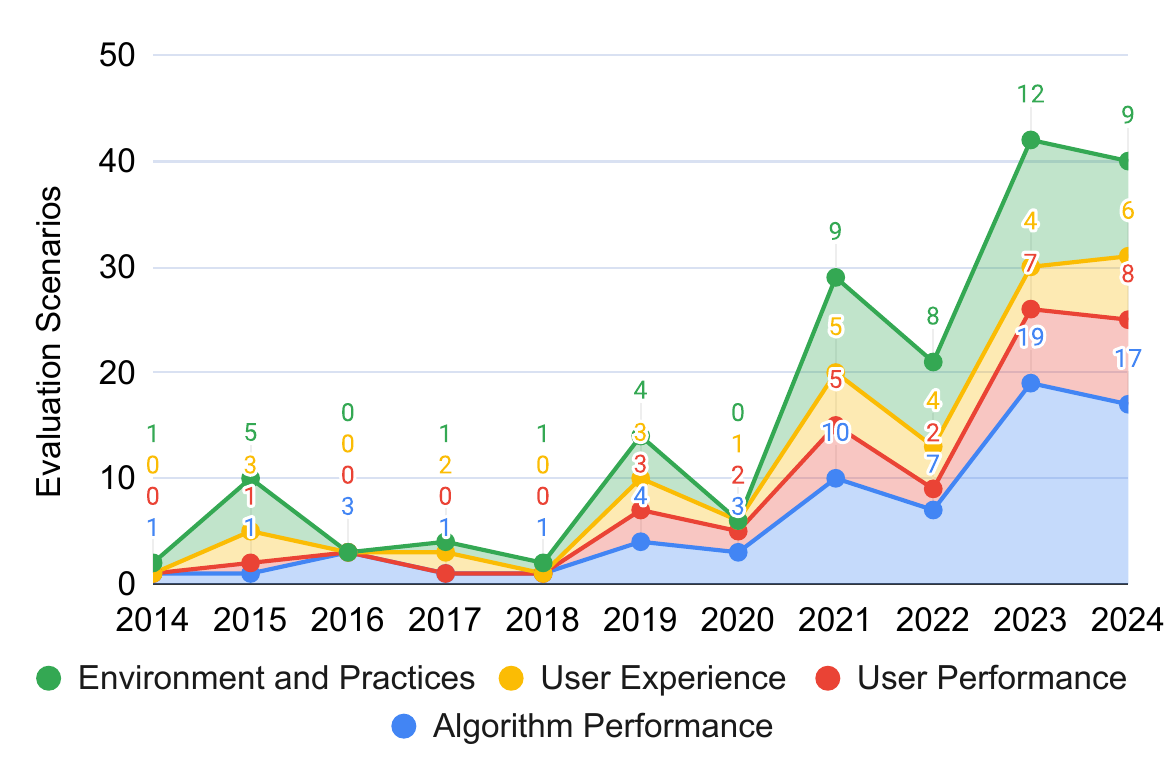}
   \caption{The evolution of the number of evaluation scenarios over the years.}
\label{fig:evaluationScenariosEvolution}
\end{figure}

\begin{figure}[ht]
  \centering
  \includegraphics[width=0.9\linewidth]{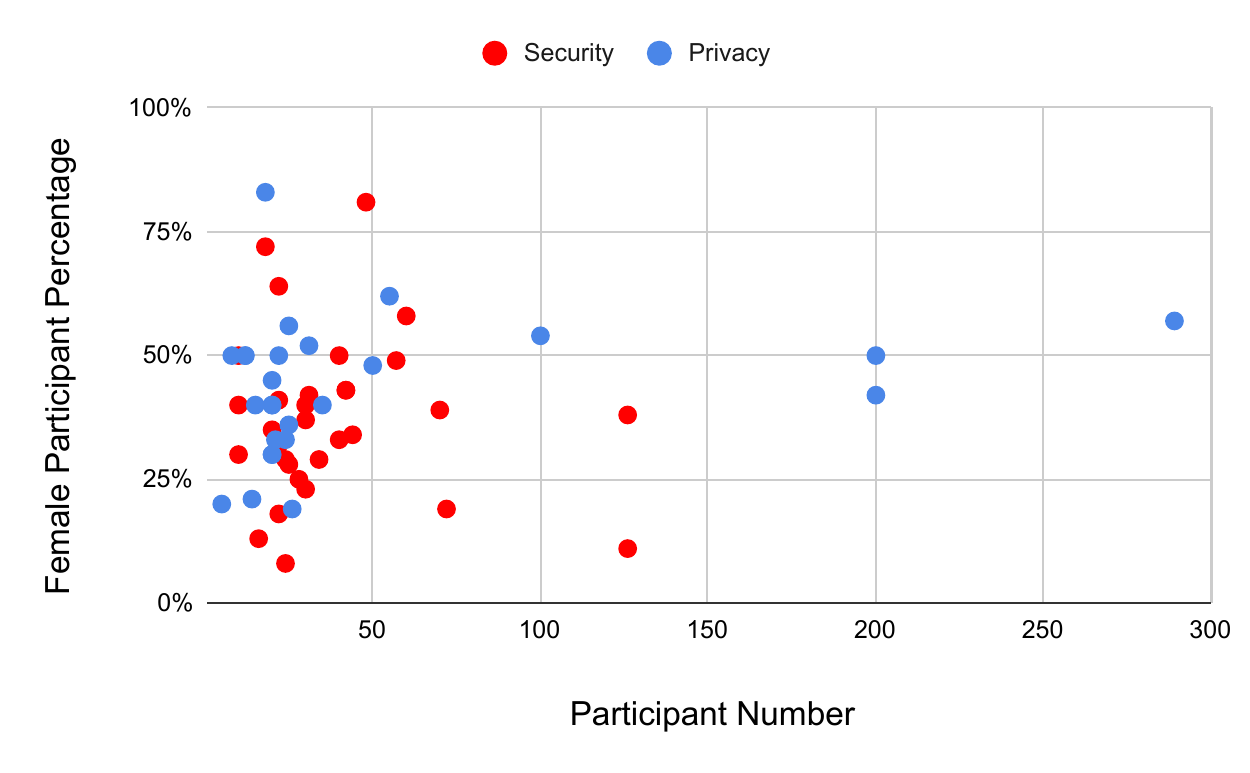}
  \caption{The sample sizes of female participants of 57 of the 80 user studies.}
  \label{fig:participants_gender_female}
\end{figure}

Figure~\ref{fig:sankeypapertypeandresearchtopic1a} presents a Sankey diagram with the distribution of the number of studies that involved the main aspects analyzed. 
In it, each column presents one of the dimensions analyzed (\ie evaluation scenarios, paper types, publication venues, security \& privacy properties, and types of reality). 
We observe that technique papers typically validate their approaches focusing on algorithm performance scenarios, whereas articles centered on evaluations mostly focus on user experience. 
The application papers present a more balanced mix of evaluation scenarios, assessing both the algorithm performance and the performance and experience of the users. 
We observed that papers in venues for human-computer interaction, such as the CHI conference, predominantly emphasize evaluations. 
In contrast, those in the TVCG journal and the IEEE VR conference are more technique-oriented. 
Articles published in security \& privacy venues, often target the USS and SP conferences, which display a balanced distribution of applications, evaluations, and techniques.
Our analysis reveals that authentication is the property most frequently examined in security-related papers.
These studies frequently incorporate virtual reality. Although a significant number of these papers appear in IEEE VR, many are distributed across various other venues.
\vspace{0cm}
\paragraph{Scenario} 
Figure~\ref{fig:evaluationScenariosEvolution} illustrates the evolution of the evaluation scenarios over the years.
Since 2018, there has been a significant increase in the evaluation of user experience, user performance, and algorithm performance scenarios. 
It is quite interesting because, in previous years, growth in these scenarios was not as high as in the past six years. 
This shift can be attributed to the increased efforts of the community and advances in technology. 
According to Stone and Chapman, the accessibility of technological developments\ins{,} such as eye tracking devices, mouse tracking devices, and similar tools\ins{,} has contributed to this increase~\citep{stone2023}. 
Moreover, the availability of software development kits (SDKs) and libraries\chg{--}{,} such as Microsoft's Mixed Reality Toolkit and \del{the }open-source solutions from Pupil Labs since 2017\ins{-}-2018\chg{--}{,} has also played an important role in driving this growth.
\vspace{0cm}
\paragraph{Quality Focus}
We note that correctness and usability are the main emphasis in metaverse evaluations.
Specifically, of the 55 studies that highlight usability, a limited number also address correctness (10), robustness (10), and time (5). 
Similarly, in the 55 studies that focus on correctness, several also consider robustness (21) and time (12). 
We think these results highlight the complexity of evaluating usability together with user performance. 
We notice that usability directly connects to user engagement and improves user compliance with privacy~\citep{metaversepaper046,metaversepaper086}
However, evaluating usability typically requires creating an intuitive interface that helps users understand system processes, ensures their continuous engagement, and encourages compliance with privacy guidelines, presenting a significant challenge for these evaluations. 
We note that half of studies (11) that relate to authentication concentrate on usability. 
The remaining studies examine both correctness and robustness and time.
For example, a study by George evaluates a method of selecting objects in three dimensions to enhance usability and security in authentication, in order to prevent shoulder surfing attacks~\citep{metaversepaper078}. 
Authentication in metaverse application is different from web-based system. 
The inclusion of additional devices, such as headsets, makes evaluating the usability of authentication methods important.
Last but not least, there are three authentication studies focused on the three aspects of quality. 
A study by Yadav focuses on designing shoulder surf-resistant PIN-based authentication mechanisms for Google Glass, using both voice-based and touch-based methods~\citep{metaversepaper031}. 
A study \chg{of}{by} Wilson analyzes privacy mechanisms for gaze data in VR, achieving re-identification accuracy as low as 14\% while maintaining high usability and task performance~\citep{metaversepaper139}.
In addition, a study by Lu provides a detailed analysis of authentication using global features from in-air handwriting signals~\citep{metaversepaper028}.
In addition, there has been limited focus on topics like policy, which address attacks such as identity theft~\citep{metaversepaper004, metaversepaper055}. 
We notice that studies that emphasize correctness and robustness instead of completion time focus on reliability and effectiveness over speed in evaluating security and privacy. 
Often, such studies are related to unobservability, which mitigates risks in identification models, typically aiming to minimize leakage and misuse risks while enhancing defenses against inference attacks.  
\begin{rmd}
\begin{itemize}
    \item 69\% of the studies analyzed involve human subjects, often focusing on aspects of user performance, such as the time needed by participants to complete security and privacy tasks and the level of accuracy they achieved.    
\end{itemize}
\end{rmd}
\paragraph{Subject} 
We collect data from user studies focusing on the metaverse and security. 
Often, these participants were involved in creating models to evaluate specific performance metrics, such as authentication or attacks. Figure~\ref{fig:participants_gender_female} presents the ratio of female participants to the total number of participants in the studies. 
Of the 80 user studies, only 57 provided gender data. 
In addition, we use two colors to visually differentiate the data, each representing security and privacy paper. 
We note that most studies (\ie 47 studies) have fewer than 50 participants and at least half are women. We found a median of 25 participants, of which 11 were female (median value). 
We observe that the sample sizes in metaverse security and privacy studies seem larger than MR/AR (median of 19 with 4 females)~\citep{merino2020evaluating}, and in HCI (median of 12)~\citep{caine2016local} in general. 
No significant differences were observed in the distribution between the security and privacy user studies, as shown in the graph.

In summary, we observed that techniques typically involve algorithm performance evaluations (\ie 39 studies), while studies centered on evaluations focus mainly on environmental practices (\ie 31 studies), user experience (\ie 15 studies), and user performance (\ie 13 studies). Applications present a more balanced mix of evaluation scenarios, assessing both algorithm performance (\ie 16 studies) and user performance and experience (\ie 14 studies). This trend highlights the diverse focus of research on metaverse technologies, emphasizing the importance of tackling both technical and user-centric challenges.
\begin{rmd}
\begin{itemize}[leftmargin=1.5em, labelsep=0.5em, itemsep=0.5em]
    \item 25 of the 114 studies implemented an open source tool such as Unity, which they made public, for instance, through an MIT license.  
    \item The sample sizes in metaverse security and privacy studies (median 25) appear larger than in MR / AR (median 19) and HCI (median 12).
\end{itemize}
\end{rmd}
\paragraph{Artifact}
Table~\ref{tab:metaversepaper4} presents details on the artifacts contained in the repositories. 
We added links in the URL column to repositories that contain source code, executable applications, or data sets. 
We confirmed that Python is the most frequent programming language, followed by C\#, JavaScript, and Java. We observe that only a few repositories have multiple stars and forks, which shows their limited relevance. 
The column \emph{license} specifies the type of license and describes how the software can be used, modified, and distributed.
Most repositories specifically have an open source license, MIT being the most frequent one. 
We note that OVRSeen has two licenses. Whereas most files are licensed under MIT, there are a few under GPLv3. 
It means that whereas most of the files of the systems allow for proprietary use and redistribution with minimal requirements (MIT), a few require that any derivative work be open-source and distributed under the same GPLv3 terms. 
LGPLv3 is a less restrictive copyleft license allowing linking with non-GPL software, and BSD-3-Clause is permissive like MIT but includes an additional non-endorsement clause. 
There are six projects without a type of license, its omission means that authors retain all rights of their source code and no one may reproduce, distribute, or create derivative works from their work, which can discourage use and contribution.
The column \emph{archive} shows whether a GitHub repository has been archived by its owner, indicating that it is no longer under active maintenance. Once archived, the repository's issues, pull requests, code, and other features become read-only. 
Contributors can only fork or star the project and cannot make direct changes unless it is unarchived.\footnote{\url{https://docs.github.com/en/repositories/archiving-a-github-repository/archiving-repositories}}. 
Interestingly, two repositories have been explicitly archived, while we notice that many others have been inactive for a long time. 
The column \emph{running} indicates the duration between the first commit and the most recent update. 
A prominent example is the \emph{HMD Eyes} project, which is very popular with 154 stars and 64 forks. 
This project showcases an open source eye-tracking platform called Pupil\footnote{\url{https://github.com/pupil-labs/pupil}} built with the Unity3D engine, specifically for Head-Mounted Displays (HMD). 
Pupil is developed by Pupil Labs\footnote{\url{https://pupil-labs.com/}}, a company focused on investigating hardware and software for eye-tracking. 
Pupil offers libraries, including an API, under the Pupil Core service, which is developed using the Python programming language. 
By integrating this platform with Unity3D, the developer aims to enhance the utility of the library, particularly in the metaverse application. 
We notice that certain projects have a notably brief duration (under two months). \texttt{TagApp} and \texttt{MetaDataStudy} are extreme examples, all of their contributions occurring in a single day.   
In addition, some studies have been found to involve multiple repositories. 
Only the repository with the most stars will be highlighted, indicating its importance. 
We confirmed that Unity is the most frequently utilized immersive framework, probably because of its active community.
\begin{table*}[ht]
\centering
\tiny
\setlength\tabcolsep{1pt}
\caption{The 25 public repositories of security and privacy tools that involve the use of the metaverse.
}
\label{tab:metaversepaper4}
\begin{tabular}
{p{1.2cm}p{2cm}p{1.3cm}p{1.1cm}p{1.7cm}p{2cm}R{0.4cm}R{0.4cm}p{1.4cm}p{0.7cm}p{1.1cm}p{1.1cm}R{1.1cm}}

  \hline
    \textbf{Paper Type} & \textbf{Ref.} & \textbf{Repo. URL} & \textbf{Artefact} & \textbf{Framework} & \textbf{Language} & \textbf{Star} & \textbf{Fork} & \textbf{License} & \textbf{Archive} & \textbf{First Commit} & \textbf{Last Update} & \textbf{Running (years)} \\\hline
    Application&\citep{metaversepaper021}&\href{https://github.com/Ethos-lab/erebus-AR_access_control}{Erebus}&AC,ED,EX&ARCore,Unity&C,C\#,Java&3&2&MIT&No&30.05.2023&21.09.2023&0.3\\
    &\citep{metaversepaper012}&\href{https://github.com/UCI-Networking-Group/OVRseen}{OVRSeen}&AC,ED,EX&Unity,Unreal&Javascript,Python&17&4&GPL-3.0,MIT&No&28.09.2021&27.10.2023&2.1\\
    &\citep{metaversepaper039}&\href{https://github.com/lehmansarahm/MAR-Security}{MAR Security}&AC,EX&-&HTML,Java&1&0&-&No&18.05.2020&03.03.2022&1.8\\
    &\citep{metaversepaper067}&\href{https://github.com/onejgordon/cec_vr}{CEC\_VR}&AC,ED,EX&Unity,Steam VR&C\#&0&0&MIT&No&14.11.2019&17.12.2020&1.1\\
    &\citep{metaversepaper111}&\href{https://github.com/arenaxr/arena-web-core}{Arena Web Core}&AC,EX&Unity&HTML, Javascript&41&28&BSD-3.0-Clause&No&17.07.2019&20.12.2024&5.4\\
    &\citep{metaversepaper079}&\href{https://github.com/pupil-labs/hmd-eyes}{HMD Eyes}&AC,EX&Unity&C\#,Python&154&64&LGPL-3.0&No&20.04.2016&22.11.2022&6.6\\
    &\citep{metaversepaper027}&\href{https://github.com/holopair-dev/HoloToolkit-Unity}{Holopair}&AC,EX&MRToolkit,Unity&C\#&1&2&MIT&No&28.01.2016&08.06.2017&1.4\\
    &\citep{metaversepaper052}&\href{https://github.com/1-max-1/nzcsc}{NZCSC}&ED&-&PHP&0&0&-&No&30.07.2020&07.07.2021&0.9\\
    &\citep{metaversepaper003}&\href{https://github.com/microsoft/prepose}{Prepose}&AC,EX&Kinect SDK&C\#&50&26&MIT&Yes&22.04.2015&26.11.2015&0.6\\
    &\citep{metaversepaper123}&\href{https://github.com/hci-pg-st23/hmd-logger}{HMD Logger}&AC,ED,EX&SteamVR&Python&3&0&-&No&08.09.2023&08.09.2023&0\\
    &\citep{metaversepaper140}&\href{https://github.com/metaguard/xror}{XROR}&EX&-&Python&5&2&BSD-3.0-Clause&No&12.04.2023&20.03.2024&0.9\\
    &\citep{metaversepaper120}&\href{https://github.com/SPICExLAB/EITPose}{EITPose}&AC,ED,EX&-&Python&10&1&MIT&No&24.01.2024&09.06.2024&0.4\\
    &\citep{metaversepaper125}&\href{https://github.com/Henrykwokkk/Meta-detector}{Meta Detector}&AC,ED,EX&-&Python&4&0&-&No&02.05.2023&30.05.2024&1.1\\
    
    Evaluation&\citep{metaversepaper022}&\href{https://github.com/MetaGuard/Identification}{MetaGuard}&AC,EX&-&Javascript,Python&13&8&MIT&No&17.02.2021&14.05.2023&2.2\\
    &\citep{metaversepaper024}&\href{https://github.com/christoftorres/Web3-Privacy}{Web3}&AC,ED,EX&-&CSS,Javascript,Python&12&1&MIT&No&23.06.2023&08.08.2023&0.1\\
    &\citep{metaversepaper060}&\href{https://github.com/ImBerkayKaplan/TagApp}{TagApp}&AC,EX&-&Assembly, C&0&0&-&No&26.01.2022&26.01.2022&0\\
    &\citep{metaversepaper042}&\href{https://github.com/MetaDataStudyAnonymized/MetaDataStudy}{MetaDataStudy}&AC,ED,EX&Unity&C\#,Python&0&0&MIT&No&30.05.2022&30.05.2022&0\\
    &\citep{metaversepaper118}&\href{https://github.com/kaiming-uw/AR_UI_Security}{AR UI Security}&AC&MRToolkit,Unity&C\#,Java&4&1&MIT&No&08.10.2023&18.11.2023&0.1\\
    
    Model&\citep{metaversepaper041}&\href{https://github.com/anonPETs2023/PETs2023}{PETs2023}&ED&Kinect SDK&Python&0&0&-&No&14.06.2022&23.08.2022&0.2\\
    &\citep{metaversepaper055}&\href{https://github.com/mvondracek/slimit}{SlimIt}&AC,ED,EX&Unity&C,Python&0&0&MIT&Yes&02.05.2011&27.06.2020&9.2\\
    
    Technique&\citep{metaversepaper086}&\href{https://github.com/metaguard/metaguard}{MetaGuard}&AC,EX&Unity&C\#&17&2&MIT&No&24.04.2022&16.08.2022&0.3\\
    &\citep{metaversepaper070}&\href{https://github.com/Terascale-All-sensing-Research-Studio/VR-Biometric-Authentication}{VR Biometric\footnote{Unaccessible since April 2025}}&AC,ED,EX&-&Python&8&2&-&No&08.06.2022&13.06.2022&0\\
    &\citep{metaversepaper072}&\href{https://github.com/BingyaoHuang/SPAA}{SPAA}&AC,ED,EX&-&Python&9&3&Custom&No&30.04.2023&13.05.2023&0\\
    &\citep{metaversepaper130}&\href{https://github.com/cschell/MoPs}{Motion}&AC,ED,EX&Unity&Python&0&0&-&No&23.08.2024&10.10.2024&0.1\\
    &\citep{metaversepaper133}&\href{https://github.com/akumar2709/Fides_AsiaCCS}{Fidel AsiaCCS}&AC,ED,EX&-&Python&0&0&-&No&16.04.2024&04.07.2024&0.2\\\hline

\end{tabular}
\end{table*}
\subsection{Potential Issues Affecting Validity}
\begin{figure}[h!]
  \centering
   \includegraphics[width=0.48\textwidth]{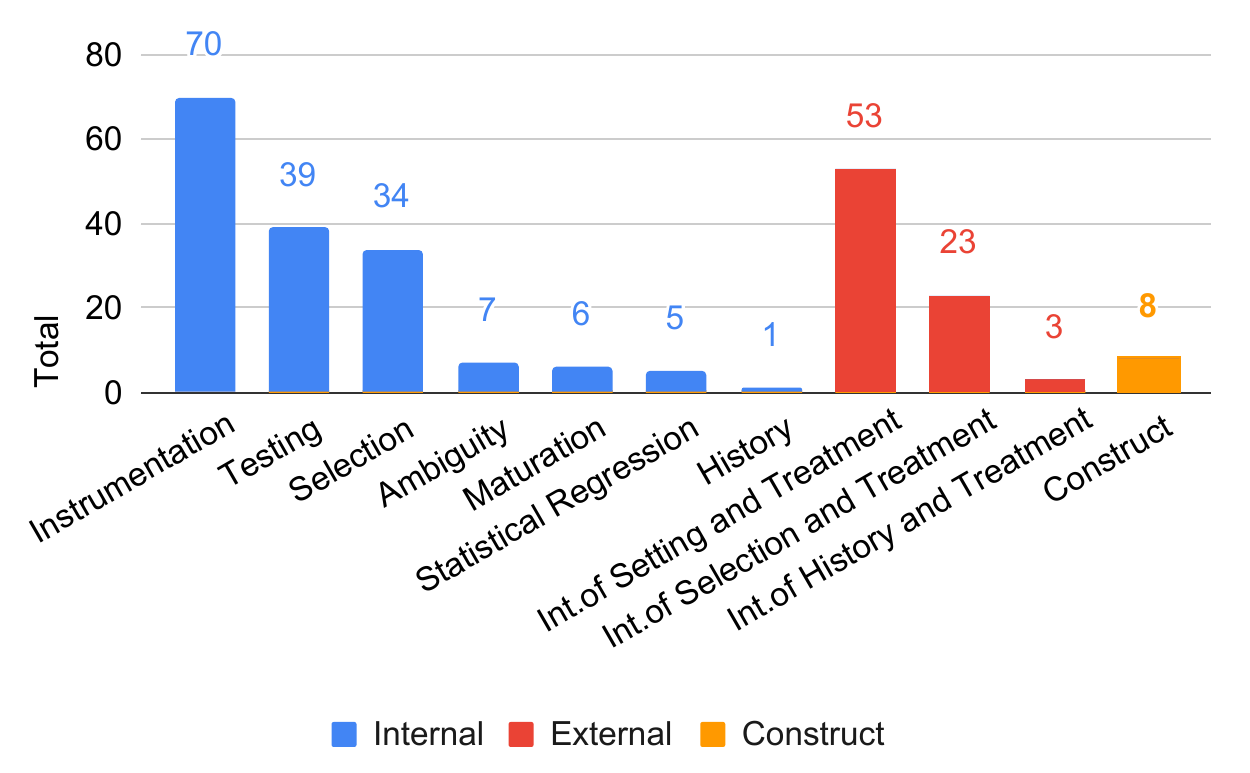}
   \caption{Classification of TTV.}
\label{fig:ttvq6}
\end{figure}

The results of the classification of the threats to validity (TTV) of the 114 studies are presented in Figure~\ref{fig:ttvq6}. In the chart, we encode threats to internal validity in blue, threats to external validity in red, and threats to construct validity in orange.
We found that threats to internal validity are the most frequent category (\ie 50\%), and did not find any studies that describe threats to conclusion validity. 

\paragraph{Internal Validity} Within internal validity threats, 28\% relate to instrumentation risks. 
Among prominent examples, we found a study in which the design assumes that a privacy-preserving eye gaze technique operates on a trusted platform, which ensures secure boot, integrity checks, and a protected open-source operating system~\citep{metaversepaper093}.
Another study introduces a type of malware that targets mixed reality headsets, operating under the assumption that it cannot modify the victim application because performing such an attack is highly complex~\citep{metaversepaper071}. Although adopting this assumption can simplify the threat model, it might not precisely reflect the genuine capabilities of attackers in the real world. 
However, they do not explicitly describe mitigation strategies.
In addition, we identified 15\% related to tests that arise when the evaluation setup influences participant behavior or when the test conditions do not accurately reflect real-world usage.
For example, a study of Li discusses a side-channel attack that occurs during the charging process of VR devices~\citep{metaversepaper137}. However, the testing setup introduced a limitation: one of the devices (MetaQuest Pro) used a charging pad that prevents it from being used while charging, unlike the other devices in the study that charged via cable.
We also found that 13\% of the articles describe instrumentation risks related to selection of settings and participants.
For example, a study \chg{of}{by} Denning selected participants from a convenience sample\del{by} inviting participants from coffee shops~\citep{metaversepaper069}. 
However, the authors acknowledge the importance of including participants from a wider range of public locations, including workplaces, playgrounds, gyms, and bars, in order to more effectively capture the diverse behaviors and social interactions that could impact the viewpoints of viewers.
Another study introduces a virtual ATM environment to examine user authentication behavior~\citep{metaversepaper107}. Although the VR setting provided a practical and cost-effective solution, it could not fully replicate the complexities of ATM interactions in the real world, such as the presence of bystanders or the pressure of being in a public space, which may have affected how participants behaved.
Interestingly, the study explicitly describes the use of realistic sound effects, which enhance the immersive experience of the virtual environment to mitigate the impact of the threat.

\paragraph{External Validity.} Threats to external validity represent 37\% of all studies and often focus on issues related to the setting and selection of participants, the latter being the most frequently addressed in 21\%.
For example, a study \chg{of}{by} Miller uses a dataset constructed from 41 right-handed users out of a total of 46 participants to train a Siamese network algorithm, \chg{which raises}{raising} concerns about the generalizability of the results of the authentication behavior obtained from the model~\citep{metaversepaper075}.
The threat is mitigated using the leave-one-out cross-validation strategy.
In other words, one out of the 41 participants is reserved for testing, while the other 40 are utilized for training. This process is repeated 41 times, with a different user being excluded each time. If the results remain consistent across these iterations, it suggests that the findings of the model could be generalized.

\paragraph{Construct Validity.} 
We found 5\% of the studies that describe threats to construct validity.
For example, Zhao's user study asks participants to provide a self-reported evaluation of their personal sense of safety~\citep{metaversepaper089}.
Although the study lacks details of the mitigation strategies used, the authors propose incorporating more objective techniques, including biometric indicators such as heart rate variability or skin conductance, in future research.

\begin{rmd}
\begin{itemize}[leftmargin=1.5em, labelsep=0.5em, itemsep=0.5em]
    \item Internal validity threats were the most prevalent, accounting for approximately 50\% of all reported validity concerns, often due to flawed assumptions and unrealistic testing setups.
    \item External validity threats appeared in 37\% of the studies, primarily due to non-representative participant samples, such as studies involving only right-handed users or limited device types.
    \item Construct and conclusion validity received minimal attention, with only 5\% of the studies addressing construct validity and none discussing threats to conclusion validity.
\end{itemize}
\end{rmd}

\paragraph{Domain-Specific Differences}
We note that validity threats are described in different ways depending on the domain (\ie security \& privacy, software engineering, and human-computer interaction).
In software engineering, threats to validity are generally addressed in a dedicated section, often located after the discussion of results and prior to the conclusions. Sometimes, threats are structured using popular classifications.
In contrast, in the security \& privacy and the human-computer interaction domains the discussion of threats to validity is scattered in the content of multiple sections (often concentrated in the discussion).
In each domain, we identified a limited number of studies specifying mitigation strategies applied to alleviate potential validity threats. However, such mentions appear more frequently in software engineering research. We think that the lack of a designated section to elaborate on validity threats in the security and privacy domain might hinder the clear articulation of mitigation measures.

\section{Threats to Validity}
\label{threats}
We discuss potential threats to the validity of our study and outline the mitigation strategies we implemented to address these threats. These threats are classified into threats to internal and external validity.
\subsection{Threats to internal validity}
\label{subsec:internalthreats}
\textit{Data collection:}
Our study relies on the accuracy of our data collection strategy. We conducted a systematic literature review, using keywords related to the metaverse and security/privacy fields to identify relevant articles. However, there may be some articles with pertinent content that we missed due to the absence of specific keywords. To address this, we iteratively refined our search queries and cross-referenced them with venues listed in the Core Ranking (A*, A, and B categories). Furthermore, we excluded duplicates and secondary studies that did not include evaluations, allowing us to maintain our focus on relevant studies.

\textit{Classification Bias:}
Human bias in the categorization and analysis of articles can impact the study's findings. To address this, two authors independently extracted and classified the data, resolving any conflicts through consensus discussions. This approach helps minimize individual biases and ensures that classifications accurately reflect the content of included studies.

\subsection{Threats to external validity}
\label{subsec:externalthreats}
\textit{Generalizability of Results:}
Our study focused on articles published in select venues that address specific immersive technologies within the metaverse. As a result, our findings may not apply to all situations or areas of the metaverse, especially those not directly included in this study. To address this limitation, we incorporated a variety of publications in the fields of software engineering, security and privacy, and human-computer interaction. In addition, our analysis is based on the content of the studies. Since these studies were published in venues with rigorous peer review processes, we are confident in their credibility and accuracy.

\section{Conclusion}
\label{conclusion}

We conducted a systematic review of security and privacy research in metaverse published between 2013 and 2024. 
We observed notable advancements in key areas such as authentication, confidentiality, and usability. However, significant gaps remain that require immediate attention from researchers and practitioners.
Limited research on back-end infrastructure and network communication protocols, coupled with the absence of scalability assessments, raises questions about the robustness of current metaverse systems in large-scale real-world deployments. Similarly, minimal attention to interoperability introduces risks for cross-platform data exchange and integration, which are pivotal to fostering a cohesive metaverse ecosystem.
The substantial focus on human participant-based evaluations underscores the importance of user-centric approaches; however, the lack of studies addressing accessibility for individuals with disabilities signifies an equity gap that could impede inclusive participation. Furthermore, regulatory and compliance considerations remain inadequately addressed, potentially leaving metaverse platforms vulnerable to privacy breaches and legal challenges.

\mg{These are interesting todos for us, which may not be relevant here. Instead please explain the implication of this study.}
\ins{To tackle these challenges, a more comprehensive and balanced research agenda is required, moving beyond isolated technical approaches toward integrated strategies that encompass usability, scalability, and ethical governance. As metaverse technologies evolve, there is an increasing need to deepen investigations into emerging areas such as AI-driven privacy-preserving mechanisms and adaptive load management that can enhance security without compromising performance or user experience. At the same time, establishing robust interoperability standards and transparent consent frameworks will be essential to build user trust across heterogeneous platforms. Taken together, these insights highlight that addressing the security and privacy challenges of the metaverse requires recognizing the deep interconnection between technical complexity and human experience, and advancing integrated interdisciplinary efforts to create immersive environments that are secure, inclusive and worthy of trust.}

\section*{CRediT authorship contribution statement}
\textbf{Argianto Rahartomo:} Writing – original draft, methodology, data curation, investigation, validation.
\textbf{Leonel Merino:} Writing – review \& editing, validation, supervision.
\textbf{Mohammad Ghafari:} Writing – review \& editing, validation, supervision.

\section*{Declaration of competing interest}
The authors declare that they have no known competing financial interests or personal relationships that could have appeared to
influence the work reported in this paper.

\section*{Acknowledgment}
Leonel Merino is funded by ANID FONDECYT Iniciaci{\'o}n Folio 11230349.

\section*{Data availability}
The full data set including the information of the studies, 
classifications, and extra figures are publicly available: \url{https://doi.org/10.5281/zenodo.15738685}.

\bibliographystyle{elsarticle-harv} 
\bibliography{main}

\newpage

\textbf{Argianto Rahartomo} a Ph.D. candidate in the Secure Software Engineering (SSE) research group of TU Clausthal, Germany. He earned a Master of Science in Internet Technologies and Information Systems at Georg-August Universität Göttingen in Germany in 2017. His research interests include security and privacy in digital environments, with a focus on applications of gamification for security, privacy challenges in the metaverse, and spectrum management.

\vspace{1em}

\textbf{Leonel Merino} is an Assistant Professor of Engineering Design in the School of Design and the School of Engineering at the Pontificia Universidad Cat\'{o}lica de Chile, Santiago, Chile. His research interests include software engineering, information visualization, virtual and augmented reality, and human-computer interaction. He received a Ph.D. in Computer Science from the University of Bern and a M.Sc. in Computer Science from the E'cole des Mines de Nantes and the Vrije Universiteit Brussel. He is a member of the Steering Committee of VISSOFT and the IEEE Computer Society. Contact him at leonel.merino@uc.cl.

\vspace{1em}

\textbf{Mohammad Ghafari} is a Professor of Software Engineering in TU Clausthal, where he leads the Secure Software Engineering (SSE) group. Mohammad’s research focus is on developing tools and techniques that facilitate secure software development. He obtained his Ph.D. in Software Engineering from Politecnico di Milano in 2015. Contact him at mohammad.ghafari@tu-clausthal.de.

\end{document}